\documentclass[aps,prd,nofootinbib,notitlepage]{revtex4}

\usepackage{graphicx,color,subfig}
\usepackage{feynmp}
\usepackage{tensor,amsfonts,amsmath,amssymb,caption}
\usepackage{hyperref}

\renewcommand{\d}{\mathrm{d}}
\renewcommand{\L}{\mathcal{L}}

\newcommand{\R}{\mathcal{R}}
\newcommand{\B}{\mathcal{B}}
\newcommand{\K}{\mathcal{K}}
\newcommand{\D}{\mathcal{D}}
\renewcommand{\tilde}[1]{\widetilde #1}
\newcommand{\order}[1]{\ensuremath{\mathcal{O}\left( #1 \right)}}
\newcommand{\ordep}[1]{\order{\epsilon^#1}}
\newcommand{\deriv}[3][d]{\frac{#1 #2}{#1 #3}}

\newcommand{\funcd}[2]{\deriv[\delta]{#1}{#2}}
\newcommand{\beq}{\begin{equation}}
\newcommand{\eeq}{\end{equation}}
\newcommand{\bes}{\begin{equation}\begin{split}}
\newcommand{\ees}{\end{equation}\end{split}}
\newcommand{\bse}{\begin{subequations}}
\newcommand{\ese}{\end{subequations}}
\newcommand{\bea}{\begin{eqnarray}}
\newcommand{\eea}{\end{eqnarray}}
\newcommand{\del}{\partial}
\newcommand{\half}{\frac{1}{2}}
\renewcommand{\vec}[1]{\mathbf{#1}}
\newcommand{\dotp}[2]{\vec #1 \cdot \vec #2}
\renewcommand{\mod}[1]{\left| #1 \right|}
\newcommand{\abs}[1]{\ensuremath{\mod{\vec #1 }}}

\newcommand{\gsim}{\raise.3ex\hbox{$>$\kern-.75em\lower1ex\hbox{$\sim$}}}
\newcommand{\lsim}{\raise.3ex\hbox{$<$\kern-.75em\lower1ex\hbox{$\sim$}}}

\newcommand{\DiffFM}{\ensuremath{\text{Diff}_{\mathcal F} (\mathcal M)}}

\newcommand{\eg}{{\it e.g.}~}
\newcommand{\ie}{{\it i.e.}~}

\begin{document}

\title{Matter in Ho\v rava-Lifshitz gravity}

\author{ Ian Kimpton} 
\email[]{ppxik1@nottingham.ac.uk}
\author{Antonio Padilla} 
\email[]{antonio.padilla@nottingham.ac.uk}

\affiliation{School of Physics and Astronomy, 
University of Nottingham, Nottingham NG7 2RD, UK} 

\date{February 21, 2013}

\begin{abstract}
We consider the role of matter in the non-projectable version of  Ho\v rava-Liftshitz gravity at both a classical and a quantum level.  At the classical level, we construct general forms of matter Lagrangians consistent with the reduced symmetry group and demonstrate that they must be reduced to their relativistic form if they are to avoid sourcing the gravitational St\"uckelberg field. At the quantum level we consider one loop corrections to the propagator for a relativistic scalar minimally coupled to gravity at tree level. We find large corrections to the light cone at low energies arising from the strength of the coupling of the scalar graviton to matter. We also find evidence that higher order time derivatives may be  generated, which is worrying  if this is to be taken seriously as a UV complete theory.
\end{abstract}

\maketitle

\tableofcontents

\section{Introduction}
Relativity and quantum mechanics underlie much of modern fundamental physics.
While both are highly successfully in their own regimes, serious problems arise combining the two.
The Standard Model excellently describes three of the fundamental forces, while gravity is left as the black sheep, only an effective theory at the quantum level.
However, there are alternative theories with the possibility of providing a quantum theory of gravity.
A recent proposal that has attracted much interest is that of Ho\v rava \cite{horava}. For reviews, see \cite{reviews}.

It has been long known that GR is perturbatively non-renormalisable \cite{'tHooft:1974bx,nonrenorm}.
This can be understood in terms of its coupling constant having negative mass dimension, $[G_N] = -2$, leading to increasingly divergent behaviour of higher order diagrams.
Several fixes have been proposed to this such as the addition of higher order derivatives to the gravitational theory.
Since these higher derivatives alter the high-energy scaling of the propagator, the coupling constant can become non-negative in the UV, rendering these theories power-counting renormalisable \cite{stelle}.
However, the presence of higher order temporal derivatives introduces ghostly pathologies, ruling them out.
Ho\v rava's proposal was to break Lorentz invariance, thereby enabling one to add higher order spatial derivatives while remaining second order in time.
The good UV behaviour is maintained, and Lorentz invariance is (hopefully) restored in the deep IR by the renormalisation group flow\footnote{However, to avoid strong coupling issues, the choice where the action exactly mimics GR is not allowed, deviations from GR must be small but non-zero.}. 

We will focus on the so called `healthy' branch of Ho\v rava's theory \cite{blas2}, which evades potential strong coupling \cite{tonyetal,blas1} and constraint algebra problems \cite{Li,Henn}, but at the expense of introducing a new scale into the theory \cite{sotpap,blas3,us}.
Our main  concern will not be  the pure gravity sector, but the coupling of Ho\v rava gravity to matter.
Gravity theories coupled with matter tend to have worse quantum behaviour than pure gravity theories \cite{'tHooft:1974bx}, and so even if pure Ho\v rava gravity is renormalisable, does it remain so when coupled to matter? Our interest here lies in one-loop corrections to the matter propagator. Such loops involving non-relativistic gravity fields will generically introduce Lorentz violation  in the matter sector.  It is  sometimes argued that supersymmetry can help suppress radiative corrections that violate Lorentz invariance \cite{Groot}, although there are doubts that a supersymmetric extension of  HL gravity can actually be found \cite{Pujolas:2011sk}. Since Lorentz Invariance is highly constrained by observation (see eg. \cite{Gagnon}) it is important to ask how much Lorentz violation will naturally occur. 
It has also been argued that the scale of Lorentz violation in the matter sector must be $M_{pl}$, not $M_\star$, due to observations of synchrotron radiation from the Crab Nebula \cite{Liberati:2012jf}.
Furthermore, in \cite{us} it was shown that Lorentz breaking terms in the matter sector source the St\"uckelberg mode in Ho\v rava gravity and can then give rise to violations of the Equivalence Principle.

This paper is made up of two main parts. In the first half of the paper we construct the general form of matter Lagrangians consistent with the reduced symmetry group of Ho\v rava gravity. For example, for a scalar field, the breaking of diffeomorphism invariance (Diff) down to foliation-preserving diffeomorphism (Diff$_{\cal F}$) allows one to add terms to the Lagrangian such as $\varphi \Delta^2 \varphi$, where  $\Delta$ is the spatial Laplacian. Assuming the time derivatives are as in the relativistic case, the general Diff$_{\cal F}$ invariant actions for a scalar field and a $U(1)$ gauge field  are given by equations (\ref{scalaradm}) and (\ref{vectoradm}) respectively.  By imposing P and T symmetry, and equivalence up to quadratic order on Minkowski space, we are able to present explicit forms for these actions (see equations (\ref{scalarF}) and (\ref{vectorterms})). The relevant actions are also written in so-called St\"uckelberg language, where diffeomorphism invariance is restored at the expense of introducing an extra field. Phenomenological difficulties can arise when the matter Lagrangian contains direct coupling to the St\"uckelberg field \cite{us} so we establish when  such couplings are absent. It turns out that they are only absent for the standard Lorentz invariant Lagrangians for both the scalar and the gauge field.

The second and most detailed part of the paper focusses on one-loop corrections to the scalar propagator. Quantum scalar fields have been studied in the context of Ho\v rava-Lifshitz gravity at the semi-classical level~\cite{arg}, whereas here we will allow gravitational fields to flow in the loops. We begin, as in  \cite{Pospelov:2010mp}, by assuming that the tree level theory is Lorentz invariant, and minimally coupled to the full spacetime metric. This is primarily because we do not want to face fine-tuning issues in the limit that gravity decouples (see \cite{Pospelov:2010mp} for discussion on this point). Since the gravity fields couple to the scalar they can flow in loops and this generically introduces Lorentz breaking as we have already suggested. Whilst there is some overlap with the analysis of  \cite{Pospelov:2010mp}, our work differs in some important ways. In particular, \cite{Pospelov:2010mp} only consider constant loop corrections to the light cone, whereas we also consider momentum dependent corrections from having generated higher order derivatives.  We also use a different method:  \cite{Pospelov:2010mp}  fix the gauge and then compute one-loop diagrams involving non-diagonal propagators. In contrast, we integrate out the constraints and work with the propagating degrees of freedom directly. While this enables us to avoid non-diagonal propagators, our method is not without some subtleties of its own. Note that we also use dimensional regularization so we only encounter logarithmic divergences. The quadratic divergences found in \cite{Pospelov:2010mp} manifest themselves as large momentum dependent corrections in our case\footnote{We thank Maxim Pospelov for pointing this out.}.

Our loop calculations reveal a number of worrying features.  The first is the large renormalisation of the light cone ( $\sim 1/\alpha \gtrsim 10^{7}$) at low energies and momentum. This follows from the fact that the scalar graviton is so strongly coupled to matter but can probably be alleviated by modifying the gravitational part of the action to include terms of the form $(D_i K_{jk})^2$.  The second issue is the generation of 
higher derivatives with respect to both space {\it and} time. The former were expected, and kick in at the Planck scale. It turns out that the UV scaling of the scalar graviton feels Planckian suppression so this is the scale that controls the higher order corrections.  The higher time derivatives, which also kick in at $M_{pl}$,  come as more of a surprise, and not a pleasant one. They suggest the presence of a new heavy ghost degree of freedom, spoiling the unitarity of the theory at high energies. Note that one finds similar behaviour in perturbative General Relativity coupled to matter although then the resulting ghost can only propagate beyond the Planck scale, outside of the regime of validity of the effective theory. In contrast, Ho\v rava gravity is intended as a UV complete theory, so if the heavy ghosts are indeed present, there is no safety net offered by an effective field theory cut-off. There is, however, some indication that this problem may be alleviated by extending the tree-level matter action to include non-relativistic terms consistent with the Lifshitz scaling of the gravity sector. This question deserves further investigation.

The rest of this paper is arranged as follows:
in section \ref{sec:nrg}, we review Ho\v rava gravity, and in particular the non-projectable version proposed by \cite{blas2}.  We then embark on the first of our two main topics in section \ref{sec:nrm} , constructing matter Lagrangians that are consistent with the reduced symmetry group of Ho\v rava gravity. In section \ref{sec:quanscal}, we consider the second of our topics, focussing on the quantum effects of matter coupled to Ho\v rava gravity and picking out the interesting features. Conclusions and discussion takes place in section \ref{sec:concl}, with some calculational details and formulae presented in an appendix.

\section{Non-relativistic gravity}\label{sec:nrg}
In Ho\v rava gravity, full diffeomorphism invariance is broken due to the special role of time in the theory, imposing that  time derivatives appear only up to second order in the action, but allowing for higher order spatial derivatives.
We are restricted to the spacetime transformations
\beq\label{difffm}
t \to \tilde t (t) \qquad x^i \to \tilde x^i (t,x).
\eeq
The result is an additional structure to that present in GR, a preferred foliation along slices of constant time.
Recall that foliations are often introduced in GR, but there they are merely a matter of convenience. The restricted transformation properties of time in Ho\v rava's theory means it is not the case here.
Two different spacetimes with slicings differing by any more than a sole function of time correspond to different physical systems.
More formally, the transformations \eqref{difffm} form the diffeomorphism subgroup which preserves the foliation, \DiffFM.

The theory is most easily formulated by making an ADM split, separating the spacetime metric $g_{\mu \nu}$ into its spatial and temporal components.
These are written as the lapse function\footnote{One can also restrict the lapse function to be solely a function of time $N(t)$. This is the projectable version of Ho\v rava theory, see \eg~\cite{projnutshell} for an overview.} $N(t,x)$, shift vector $N^i (t,x)$ and spatial metric on the slice $\gamma_{ij} (t,x)$,
\beq
\d s^2 = g_{\mu \nu} \d x^\mu \d x^\nu = - N^2 \d t^2 + \gamma_{ij} (\d x^i + N^i \d t) (\d x^j + N^j \d t).
\eeq
Under \eqref{difffm}, these fields transform as
\begin{subequations}\label{fpdifffields}
\begin{align}
\delta \gamma_{ij} &\to \delta \gamma_{ij} + 2 D_{(i} \zeta_{j)} + f \dot \gamma_{ij} \label{fpdiffN}\\
\delta N_i &\to \delta N_i + \del_i \left( \zeta^j N_j \right) - 2 \zeta^j D_{[i} N_{j]} + \dot \zeta^j \gamma_{ij} + \dot f N_i + f \dot N_i \\
\delta N &\to \delta N + \zeta^j \del_j N + \dot f N + f \dot N,
\end{align}
\end{subequations}
where $D_i$ is the covariant derivative associated with $\gamma_{ij}$ and dots denote $\deriv{}{t}$.

We now construct a gravitational action consistent with \eqref{difffm}.
We impose that the theory have no higher than second order time derivatives, to avoid the associated ghostly instabilities. However, the loss of Lorentz invariance means that one is permitted to add higher order spatial derivatives, with the number of these denoted  by $2z$.
The additional spatial derivatives cause the propagator of the graviton to fall off faster in the UV than occurs in GR, and it is therefore argued that the theory will be power-counting renormalisable.
Clearly, time and space scale anisotropically in the UV, resulting in scaling dimensions (in $D+1$ dimensional spacetime) in the UV of
\beq
[t] = - z \qquad [x^i] = -1 \qquad [G] = z-D.
\eeq
Obviously, $z = 1$ in a relativistic theory.
We restrict ourselves to the case of $D=3$ dimensions, and so for a power-counting theory we require $z \geq 3$ (since then $[G] \geq 0$). Here, as is common in most QFTs, we consider the marginal case $z=3$.

The kinetic piece of the action is constructed from the extrinsic curvature of the spatial slices,
\beq
K_{ij} = \frac 1 {2N} \left( \dot \gamma_{ij} - 2 D_{(i} N_{j)} \right).
\eeq
Denoting the potential (containing our spatial derivatives) by $S_V$, the gravitational action can be written
\beq\label{Sgrav}
S_{grav} = {M_{pl}^2} \int d t d^3 x \sqrt \gamma N \left( K_{ij} K^{ij} - \lambda K^2 \right) + S_V,
\eeq
where $M_{pl}$ is the Planck mass.
This kinetic piece differs from GR by the introduction of the $\lambda$ parameter.
This takes the value $1$ in GR, but the reduced symmetries of Ho\v rava theory mean that this number is not fixed in Ho\v rava gravity and indeed it is expected to run under the RG flow.

The gravitational potential is built from objects satisfying the \DiffFM~symmetry, up to sixth order in spatial derivatives.
The exhaustive list of building blocks is the (inverse) metric $\gamma^{ij}$, the Ricci tensor of a slice $R_{ij}$ and $a_i \equiv \del_i \log N$ \cite{blas2}, the acceleration of  spatial slices through the spacetime. We write this piece of the action as
\beq
S_V =  {M_{pl}^2} \int dt d^3 x \sqrt \gamma N \left( R + \alpha a_i a^i + \frac 1 {M_{pl}^2} V_4 + \frac 1 {M_{pl}^4} V_6 \right),
\eeq
where the four derivative $V_4$ and six derivative $V_6$ terms are given by
\begin{subequations}\label{ADMsplitgravpot}
\begin{align}
V_4 &= A_1 R_{ij} R^{ij} + A_2 R^2 + A_3 R D_i a^i + A_4 (D_i a^i)^2 \\
V_6 &= B_1 (D_i R_{jk})^2 + B_2 (D_i R)^2 + B_3 \triangle R D_i a^i + B_4 a^i \triangle^2 a_i
\end{align}
\end{subequations}
where $\triangle \equiv D_i D^i$ and we only include terms which are inequivalent at quadratic order around a Minkowski background\footnote{Some of our expansions will go to higher order, but including just the terms here will capture all the relevant physics.}.
To ensure the absence of strong coupling in the theory (needed to ensure that the theory remains perturbative and so our power counting argument can hold), one needs to introduce a hierarchy of scales by making the $B$s large \cite{us,blas2}. For definiteness we assume $A_i \sim \order 1$ and $B_i \sim 1/\alpha$ \cite{blas2}. Constraints for $\lambda$ and $\alpha$ give roughly $\mod{1 - \lambda} \sim \alpha \lesssim 10^{-7}$ \cite{blas3}, or $B \gtrsim 10^{7}$.
It turns out that  two  new scales  are introduced, $M_\star \sim  \sqrt \alpha M_{pl}$ and $M_h \sim \alpha^{1/4} M_{pl}$ \cite{blas3}. Putting all these pieces together, one obtains the full action for Ho\v rava gravity.

As we will see in section \ref{sec:nrm}, it is often more illuminating to write Ho\v rava gravity in a form using covariant 4D spacetime tensors.
The St\"uckelberg trick \cite{stucky} allows one to artificially restore (full) gauge invariance at the expense of an additional scalar field. The difference between Ho\v rava gravity and a fully diffeomorphism invariant theory like GR boils down to the foliation structure and the $t = constant$ hypersurfaces. We therefore introduce the St\"uckelberg field $\phi = \phi(x^\mu)$ and redefine the foliation \cite{germani,blas2}
\beq
 t = constant \quad \to \quad \phi  = constant,
\eeq
where $\phi$ is some scalar function of the spacetime coordinates. The choice $\phi = t$ obviously will give us back the original formulation. We now introduce the unit normal to the hypersurfaces,
\beq
u^\mu = \frac{\nabla^\mu \phi} X \qquad X = \sqrt{- \nabla_\mu \phi \nabla^\mu \phi},
\eeq
where $\nabla$ is the 4D covariant derivative associated with the 4D metric $g_{\mu \nu}$. The induced metric on the spatial slices can then be promoted to a 4D tensor,
\beq
\gamma_{i j} \to \gamma_{\mu \nu} = g_{\mu \nu} + u_\mu u_\nu,
\eeq
which is a projector onto the spacelike submanifold defined by the timelike normal $u_\mu$. One can now promote all the other relevant quantities to tensors
\begin{subequations}\label{promquan}
\begin{align}
K_{ij} &\to \K_{\mu \nu} = \half \pounds_u \gamma_{\mu \nu} = \gamma^{\alpha}_{(\mu} \gamma^{\beta}_{\nu)} \nabla_\alpha u_\beta \\
R_{i j} &\to \R_{\mu \nu} = \gamma_\mu^\alpha \gamma_\nu^\beta \gamma^{\rho \sigma} R^{(4)}_{\rho \alpha \sigma \beta} + \K_{\mu \alpha} \K_\nu^\alpha - \K \K_{\mu \nu} \\
a^i &\to a^\mu = u^\nu \nabla_\nu u^\mu,
\end{align}
\end{subequations}
and it is useful to introduce the spatially projected covariant derivative,
\beq\label{spcd}
D_i X^{j_1 \cdots j_n} \to \D_\mu X^{\alpha_1 \cdots \alpha_n} = \gamma_\mu^\nu \gamma^{\alpha_1}_{\beta_1} \cdots \gamma^{\alpha_n}_{\beta_n} \nabla_\nu X^{\beta_1 \cdots \beta_n}.
\eeq
We now write the action \eqref{Sgrav} in a covariant form,
\beq
S_{grav} = M_{pl}^2 \int d^4 x \sqrt{-g} R^{(4)} + \triangle S_K + \triangle S_V,
\eeq
where
\begin{subequations}
\begin{align}
\triangle S_K &= (1 - \lambda) M_{pl}^2 \int d^4 x \sqrt{-g} \K^2 \\
\triangle S_V &= M_{pl}^2 \int d^4 x \sqrt{-g} \left[ \alpha a^\mu a_\mu + \frac{V_4}{M_{pl}^2} + \frac{V_6}{M_{pl}^4} \right],
\end{align}
\end{subequations}
are the additional kinetic and potential pieces in the theory (as compared with GR), and with $V_4$ and $V_6$ given by the covariantised versions of \eqref{ADMsplitgravpot}.

This formulation has the advantage of easy comparison with GR, and it can also be used to calculate how matter should couple to gravity in this theory. It is easy to show (see \cite{us}) that
\beq\label{mattercoupling}
\gamma_{\alpha \nu} \nabla_\mu T^{\mu \nu} = 0 \qquad \frac 1 {\sqrt{-g}} \funcd{S_m}{\phi} = - \frac 1 X \nabla_\mu \phi \nabla_\nu T^{\mu \nu},
\eeq
where $T^{\mu \nu} = \frac 2 {\sqrt{-g}} \funcd{S_m}{g_{\mu \nu}}$ is the energy momentum tensor derived from the matter action $S_m$. Note that matter sources the St\"uckelberg field directly when there is some violation of energy-momentum conservation. Energy-momentum conservation is linked to diffeomorphism invariance which is absent here, so some violation can occur. Such violations can, in principle, lead to violations of the Equivalence Principle \cite{us}. 
\section{Non-relativistic matter}\label{sec:nrm}
We now consider the first main topic of this paper: what is the general form of matter Lagrangians consistent with the {\DiffFM} symmetry of Ho\v rava gravity?  Lorentz invariant matter actions, minimally coupled to the spacetime metric are expected to receive quantum corrections via gravity loops that spoil the Lorentz invariance. Indeed, in the second main topic of this paper we will see explicitly how  this is the case.  For now, however, let us try to formulate the relevant actions for a scalar and a $U(1)$ gauge field, consistent with the foliation of spacetime.   Of course, these will differ from the standard Lorentz invariant actions because extra terms are allowed due to the reduced symmetry. The relevant actions will be considered in both the ADM and St\"uckelberg formalisms, as in the case of the gravitational action. The analysis here is similar to that of \cite{KirKof}, but we also consider the possible effect of $a_i$ terms  here.  Note that in keeping with the philosophy of Ho\v rava gravity we only consider generalisations to the potential and assume the time derivatives enter as in the relativistic theory. This ensures the absence of ghostly instabilities, but as we will see later on, it is not guaranteed at one loop.

\subsection{Scalar field}\label{sec:nrm-scal}
For a scalar field $\varphi$, the generic action with a {\DiffFM} invariant potential can be written in the ADM formalism as
\beq\label{scalaradm}
S_\varphi = \int d t d^3 x \sqrt{\gamma} N \left[ \frac{1}{2N^2} (\dot \varphi - N^i \del_i \varphi)^2 - \half \gamma^{ij} \del_i \varphi \del_j \varphi - V(\varphi)   - F[\varphi, D_i , R_{ij}, a_i, \gamma^{ij}] \right].
\eeq
The symmetries of the Ho\v rava framework permit additional terms in the theory relative to GR, which $F$ controls ($F=0$ is the usual minimally-coupled diffeomorphism invariant action). As in the gravitational sector, we will only consider terms up to scaling dimension 6.
These terms can be constructed from $\varphi$, $D_i$, $R_{ij}$, $a_i$ and $\gamma^{ij}$ and make up a general $F$.
We will enforce P and T symmetries, neglect purely gravitational terms, and only consider inequivalent terms at quadratic order on Minkowski.
This results in
\beq
\begin{split}
\label{scalarF}
F = & \alpha_1 \varphi D^i a_i + \alpha_2 \varphi \triangle \varphi + \alpha_3 R \varphi + \frac{\beta_1}{M_{pl}^2} D^i a_i  \triangle \varphi + \frac{\beta_2}{M_{pl}^2} \varphi \triangle^2 \varphi + \frac{\beta_3}{M_{pl}^2} R \triangle \varphi \\ & + \frac{\gamma_1}{M_{pl}^4} D^i a_i  \triangle^2 \varphi + \frac{\gamma_2}{M_{pl}^4} \varphi \triangle^3 \varphi + \frac{\gamma_3}{M_{pl}^4} R \triangle^2 \varphi ,
\end{split}
\eeq
where $\triangle \equiv D_i D^i$. This list is exhaustive given our above restrictions, since terms such $R_{ij} D^i D^j \varphi$ are equivalent to other terms via the Bianchi identity, and others such as $R D_i \varphi D^i \varphi$ are ignored since they vanish at quadratic order on Minkowski space.

These expressions can also be re-written using the St\"uckelberg formulation. Again, as with gravity, the action is simpler in this formalism, and can be written
\beq
S_\varphi = \int d^4 x \sqrt{-g} \left[ - \half g^{\mu \nu} \nabla_\mu \varphi \nabla_\nu \varphi - V(\varphi) - F \right].
\eeq
where $F$ is now
\beq
\begin{split}
\label{scalarFstuck}
F = & \alpha_1 \varphi \D_\mu a^\mu  + \alpha_2 \varphi \D^\mu \D_\mu \varphi + \alpha_3 \R \varphi
 + \frac{\beta_1}{M_{pl}^2} \D_\mu a^\mu  \D^\nu \D_\nu \varphi + \frac{\beta_2}{M_{pl}^2} \varphi (\D^\mu \D_\mu)^2 \varphi + \frac{\beta_3}{M_{pl}^2} \R \D^\mu \D_\mu \varphi \\
  & + \frac{\gamma_1}{M_{pl}^4} \D_\mu a^\mu  (\D^\nu \D_\nu)^2 \varphi + \frac{\gamma_2}{M_{pl}^4} \varphi  (\D^\nu \D_\nu)^3 \varphi + \frac{\gamma_3}{M_{pl}^4} \R  (\D^\nu \D_\nu)^2 \varphi .
\end{split}
\eeq
In this formalism, the recovery of the usual minimally coupled scalar field in the case $F=0$ is clearer.

It is now possible to ask in what way the action must be constructed to avoid any coupling to the St\"uckelberg field (which one may want to avoid for reasons of \eg~Equivalence Principle \cite{us} or Lorentz violation).
In fact, it is easy to show that the only combination of terms which does not couple matter to the St\"uckelberg field is the usual Lorentz invariant action with $F \equiv 0$. 

To see this, we set  $\frac{\delta}{\delta \phi} \int d^4 x \sqrt{-g} F=0$. Strictly speaking we only require this to vanish on-shell,  but since $\varphi$ can be coupled to a source independently of its coupling to the St\"uckelberg field, it is clear that we need to impose $\frac{\delta}{\delta \phi} \int d^4 x \sqrt{-g} F=0$ off-shell in order to guarantee $\frac{\delta S_\varphi}{\delta \phi}=0$ in all cases.  Now, because the necessary  cancellation can only occur between terms with the same power of $M_{pl}$ and the same number of $\varphi$'s, it immediately follows that $\alpha_2=\beta_2=\gamma_2=0$, and that 
\begin{eqnarray}
\frac{\delta}{\delta \phi} \int d^4 x \sqrt{-g} \left[\alpha_1 \varphi \D_\mu a^\mu + \alpha_3 \R \varphi\right] &=& 0 \label{alpha13} \\
\frac{\delta}{\delta \phi} \int d^4 x \sqrt{-g} \left[\beta_1\D_\mu a^\mu  \D^\nu \D_\nu \varphi + \beta_3 \R \D^\mu \D_\mu \varphi \right] &=& 0 \label{beta13}\\
\frac{\delta}{\delta \phi} \int d^4 x \sqrt{-g}  \left[\gamma_1 \D_\mu a^\mu  (\D^\nu \D_\nu)^2 \varphi+\gamma_3 \R  (\D^\nu \D_\nu)^2 \varphi \right]  &=& 0 \label{gamma13}
\end{eqnarray}
Consider equation (\ref{alpha13}). Introducing $\tilde \varphi=\frac{\sqrt{-g}}{\sqrt{\gamma} }\varphi$, this implies that
\beq
\alpha_3 \sqrt{\gamma} \left[\R_{\mu\nu}-\frac{1}{2} \R \gamma_{\mu\nu}\right] \tilde \varphi +\text{terms with derivatives of $\tilde \varphi$}=0
\eeq
Since this should be true for any $\varphi$ and $\gamma_{\mu\nu}$, we conclude that $\alpha_3=0$. Furthermore, since $\frac{\delta}{\delta \phi} \int d^4 x \sqrt{-g}\varphi \D_\mu a^\mu  \neq 0$, in general, it also follows that $\alpha_1=0$. Similar arguments can be applied to equations (\ref{beta13}) and (\ref{gamma13}) to conclude that $\beta_1=\beta_3=0$, and $\gamma_1=\gamma_3=0$. It now follows that $F \equiv 0$, as previously stated.

\subsection{Gauge field}
We also consider a vector field $A^\mu$ invariant under a $U(1)$ gauge symmetry (see also \cite{elecalahorava}).
Our analysis will run along much the same lines as for the scalar field. The general action, invariant under {\DiffFM} can be written in terms of ADM variables as,
\beq\label{vectoradm}
S_A = \frac{1}{4} \int dt d^3 x \sqrt{\gamma} N \left[ \frac{2}{N^2} \gamma^{ij} (F_{0i} - F_{ki}N^k) (F_{0j} - F_{lj}N^l) - F_{ij} F_{kl} \gamma^{ik} \gamma^{jl} - G \right],
\eeq
where $G=0$ in the familiar relativistic case and $F_{\mu \nu} \equiv \del_\mu A_\nu - \del_\nu A_\mu$.
There is an additional constraint on the possible terms in $G$ since we are demanding that the theory remain gauge invariant with respect to the $U(1)$.
In order to add higher spatial derivatives, it is convenient to write the higher order terms containing the vector field in terms of the magnetic field,
\beq
B^i = \half \frac{{\varepsilon}^{ijk}}{\sqrt \gamma} F_{jk},
\eeq
where $\varepsilon^{ijk}$ is the Levi-Civita symbol. The magnetic field corresponds to the only gauge invariant way that higher-order spatial derivatives of $A$ can enter\footnote{The electric field corresponds to time derivatives, so  additional electric field terms result in higher order time derivatives.}.
$G$ can be built therefore from $B^i, D_i, R_{ij}, a_i, \gamma^{ij}$.
Assuming P and T symmetry again, the terms inequivalent at quadratic level on Minkowski and up to scaling dimension 6 are
\beq
\begin{split}\label{vectorterms}
G &= \alpha_1 a_i B^i + \alpha_2 B_i B^i + \frac{\beta_1}{M_{pl}^2} a_i \triangle B^i  + \frac{\beta_2}{M_{pl}^2} B_i \triangle B^i + \frac{\beta_3}{M_{pl}^2} (D_i B^i)^2 + \frac{\beta_4}{M_{pl}^2} R D_i B^i \\
& + \frac{\gamma_1}{M_{pl}^4} B_i \triangle^2 a^i + \frac{\gamma_2}{M_{pl}^4} B_i \triangle^2 B^i + \frac{\gamma_3}{M_{pl}^4} (D_i D_j B^j)^2 + \frac{\gamma_4}{M_{pl}^4} R \triangle D_i B^i .
\end{split}
\eeq

In order to also write the vector field action in the St\"uckelberg approach, we need a four-vector expression for $B_i$. The appropriate expression is
\beq
\B^\mu = \half \frac{\epsilon^{\nu \mu \rho \sigma}}{\sqrt{-g}} F_{\rho \sigma} u_\nu.
\eeq
Note that the St\"uckelberg coupling comes in to this term directly  via the normal term $u_\nu$. Proceeding as before, our action is now the familiar
\beq
S_A = \int \d^4 x \sqrt{-g} \left[ - \frac 1 4 F^{\mu \nu} F_{\mu \nu} - G \right].
\eeq
By making the substitutions $B_i \to \B_\mu$, $D_i \to \D_\mu$, $a_i \to a_\mu$, $R_{ij} \to \R_{\mu \nu}$ and $\gamma^{ij} \to \gamma^{\mu \nu}$ into \eqref{vectorterms}, the general expression for $G$ can be written in this formalism.
In this case, the St\"uckelberg field couples through the projection operator to the matter field, as well as to the magnetic field through the normal. As with the scalar field, we note that the only way to prevent a coupling between the matter field and the St\"uckelberg field is to set $G \equiv 0$ in our action.

\section{Quantum corrected matter}\label{sec:quanscal}
We now turn to the second major topic of this paper: quantum corrections to relativistic matter Lagrangians. In particular we will consider one loop corrections to the relativistic scalar field action with mass $m$ and a $\varphi^4$ interaction, 
\beq \label{stree}
S^{tree}_\varphi=\int dt d^3 \vec x \sqrt{-g} \left[ -\frac{1}{2} g^{\mu\nu} \nabla_\mu \varphi \nabla_\nu \varphi-\frac{1}{2} m^2 \varphi^2 -\frac{\mu}{4!} \varphi^4\right]
\eeq

Since we are interested in the role played by the Lorentz violating gravity sector, we will include loops of gravity fields, and expect these to induce some Lorentz violation in the scalar field theory. We will concentrate on corrections to the scalar field propagator, including the contribution from higher order spatial derivatives,   in contrast to \cite{Pospelov:2010mp} who only considered constant corrections to the light cone. Our method also differs to that in \cite{Pospelov:2010mp}: they fix the gauge  and work with non-diagonal propagators, whereas we integrate out the constraints and work directly with the dynamical degrees of freedom.  This method has the advantage of allowing us to work with diagonal propagators, but is not without its subtleties as we will illustrate by means of a toy model in the next section. Note that the effective action we obtain is consistent in that the resulting classical dynamics is independent of when we impose the constraints \ie~before we compute the equations of motion, or after.

\subsection{Toy model}\label{sec:intoutex}
As a warm up to the main event we consider the following toy model of a dynamical scalar, $\phi$, coupled to a non-dynamical scalar, $A$.
\beq\label{intoutexlag}
\L = - \half (\del_\mu \phi)^2 - \half A \triangle A - \half m^2 A^2 + \lambda \phi^2 A,
\eeq
where $\triangle \equiv \del_i \del^i$.  Our interest lies the one-loop corrections to the propagator for $\phi$. We can compute this in two ways:  directly from the Lagrangian (\ref{intoutexlag}) by defining a propagator for both $\phi$ and $A$; or by integrating out the non-dynamical field, $A$, and only working with the dynamical degree of freedom, $\phi$. We will compare the two methods, beginning with the former.

The Lagrangian (\ref{intoutexlag})  gives rise to the following field equations
\bea
&&\frac{\delta}{\delta \phi} \int d^4 x \L=\Box \phi+2\lambda \phi A=0 \label{peq}\\
 &&\frac{\delta}{\delta A} \int d^4 x \L=-(\Delta +m^2) A+\lambda \phi^2=0 \label{conA}
\eea
and a set of Feynman rules shown in Figure \ref{fig:IOE1FR}. At one loop the correction to the $\phi$ propagator comes from the  Feynman diagrams shown in Figure \ref{fig:IOE1CP}. 
\begin{figure}[h]
	    \vspace{9pt}
  \subfloat{\begin{minipage}{0.3\linewidth}
    \begin{center}
    \unitlength = 1mm
    \begin{fmffile}{IOE1FR1}
  \begin{fmfchar*}(25,25)
  \fmfleft{i1}  \fmflabel{$\phi$}{i1}
  \fmf{plain,label=$p$} {i1,o1}
  \fmfright{o1}  \fmflabel{$\phi$}{o1}
\end{fmfchar*}
\end{fmffile} \vspace{-20pt} $$ \frac{-\mathrm i}{p^2}$$
    \end{center}
  \end{minipage}}
  \subfloat{\begin{minipage}{0.3\linewidth}
    \begin{center}
    \unitlength = 1mm
    \begin{fmffile}{IOE1FR2}
  \begin{fmfchar*}(25,25)
  \fmfleft{i1} \fmflabel{$A$}{i1}
  \fmf{boson,label=$p$} {i1,o1}
  \fmfright{o1} \fmflabel{$A$}{o1}
\end{fmfchar*}
\end{fmffile} \vspace{-20pt} $$ \frac{-\mathrm i}{-\abs p^2 + m^2}$$
    \end{center}
  \end{minipage}}
  \subfloat{\begin{minipage}{0.3\linewidth}
    \begin{center}
    \unitlength = 1mm
    \begin{fmffile}{IOE1FR3}
  \begin{fmfchar*}(25,25)
  \fmfleft{i1,i2} \fmflabel{$\phi$}{i1} \fmflabel{$\phi$}{i2}
  \fmf{plain} {i1,v1}
  \fmf{plain} {i2,v1}
  \fmf{boson} {o1,v1}
  \fmfright{o1} \fmflabel{$A$}{o1}
  \fmfdot{v1}
\end{fmfchar*}
\end{fmffile} \vspace{-20pt} $$ 2 i \lambda$$
\vspace{9pt}
    \end{center}
  \end{minipage}}
\caption{The Feynman rules for the Lagrangian  (\ref{intoutexlag}) }
\label{fig:IOE1FR}
	    \vspace{9pt}
\end{figure}
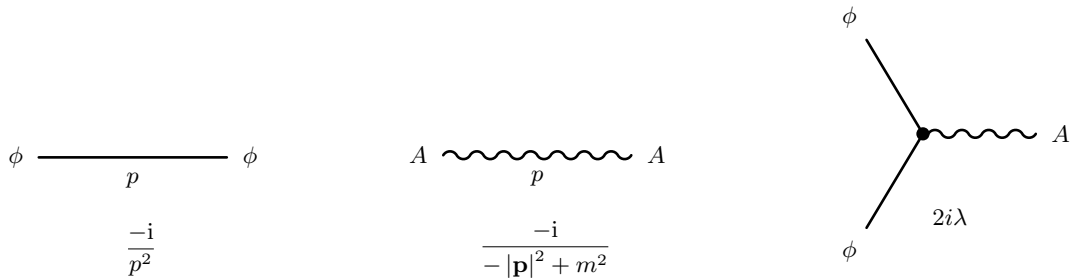
\begin{figure}[h]
	    \vspace{9pt}
  \subfloat[1PI contribution]{\label{fig:IOE1CPa}\begin{minipage}{0.3\linewidth}
    \begin{center}
    \unitlength = 1mm
    \begin{fmffile}{IOE1CP1}
  \begin{fmfchar*}(25,25)
\fmfleft{i1}
\fmfright{o1}
\fmf{plain,tension=3}{i1,v1}
\fmf{plain,tension=3}{v2,o1}
\fmf{plain,tension=3}{v1,v2}
\fmf{boson,left,tension=-3}{v1,v2}
\fmf{phantom}{v1,v2}
\fmfdot{v1,v2}
\end{fmfchar*}
\end{fmffile} \vspace{-20pt}
    \end{center}
  \end{minipage}}
  \subfloat[Tadpole contribution]{\label{fig:IOE1CPb}\begin{minipage}{0.3\linewidth}
    \begin{center}
    \unitlength = 1mm
    \begin{fmffile}{IOE1CP2}
  \begin{fmfchar*}(25,25)
\fmfright{o1} \fmfleft{i1} \fmftop{v2}
\fmf{plain,tension=5}{i1,v1}
\fmf{plain,tension=5}{v1,o1}
\fmf{photon,tension=0}{v1,v2}
\fmf{plain}{v2,v2}
\end{fmfchar*}
\end{fmffile} \vspace{-20pt}
    \end{center}
  \end{minipage}}
\caption{One-loop diagrams for the $\phi$ propagator}
\label{fig:IOE1CP}
	    \vspace{9pt}
\end{figure}
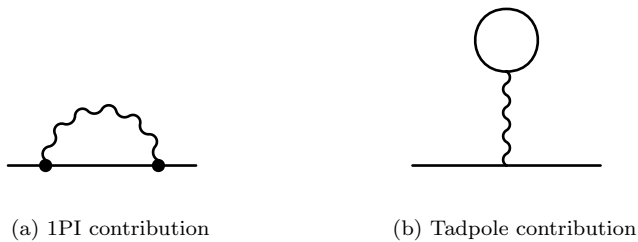
The one-loop correction contains a 1PI contribution and a tadpole contribution. In contrast to QED, here the tadpole contribution need not vanish. Indeed, from Figure \ref{fig:IOE1CPa}, we find the 1PI contribution to be 
\beq\label{IOEA}
\textrm{1PI}=\left( \frac{- \mathrm i}{k^2} \right)^2 (2 \mathrm i \lambda)^2 \int \frac{d^4 p}{(2 \pi)^4} \frac{ - \mathrm i}{p^2} \frac{ - \mathrm i}{- \mod{\vec k - \vec p}^2 + m^2} = - \frac{4 \lambda^2}{k^4} \int \frac{d^4 p}{(2 \pi)^4} \frac 1 {p^2 \left( - \mod{ \vec k-\vec p }^2 + m^2 \right)} .
\eeq
whereas from Figure \ref{fig:IOE1CPb}, we find the tadpole contribution to be 
\beq\label{IOEB}
\textrm{tadpole}=\half \left( \frac{- \mathrm i}{k^2} \right)^2 (2 \mathrm i \lambda)^2 \int \frac{d^4 p}{(2 \pi)^4} \frac{ - \mathrm i}{p^2} \frac{ - \mathrm i}{ m^2} = - \frac{2 \lambda^2}{k^4} \int \frac{d^4 p}{(2 \pi)^4} \frac 1 {p^2 m^2}.
\eeq
Now a non-vanishing tadpole is the same as saying that the vev of the field $A$ is non-vanishing. One could add a counterterm to the Lagrangian of the form $\Delta \L=(\textrm{constant}) A$ in order to eliminate this, and therefore eliminate the tadpole. The spirit of this discussion is particularly relevant for matter loops in Ho\v rava gravity to be studied in subsequent sections. The point is that in Ho\v rava gravity  matter loops also endow the gravitational fields with a non-trivial vev because the theory offers no solution to the cosmological constant problem. By inserting a bare cosmological constant into the action as a counterterm one can eliminate the vevs of those fields. In the subsequent section we will assume that this has been done by neglecting the tadpole contribution from the relevant diagrams. 

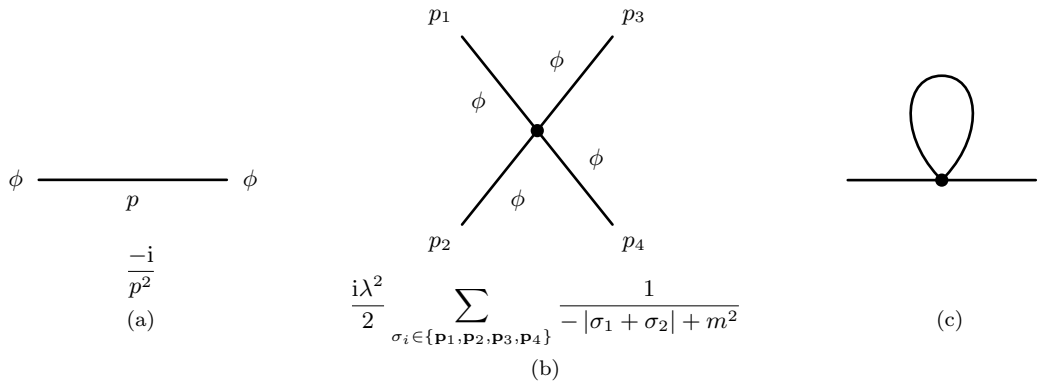
\begin{figure}[h]
	    \vspace{9pt}
  \subfloat[]{\label{IOE2a}\begin{minipage}{0.3\linewidth}
    \begin{center}
    \unitlength = 1mm
    \begin{fmffile}{IOE21}
  \begin{fmfchar*}(25,25)
  \fmfleft{i1}  \fmflabel{$\phi$}{i1}
  \fmf{plain,label=$p$} {i1,o1}
  \fmfright{o1}  \fmflabel{$\phi$}{o1}
\end{fmfchar*}
\end{fmffile} \vspace{-20pt} $$ \frac{-\mathrm i}{p^2}$$
    \end{center}
  \end{minipage}}
  \subfloat[]{\label{IOE2b}\begin{minipage}{0.3\linewidth}
    \begin{center}
    \unitlength = 1mm
    \begin{fmffile}{IOE22}
  \begin{fmfchar*}(25,25)
  \fmfleft{i1,i2} \fmflabel{$p_2$}{i1} \fmflabel{$p_1$}{i2}
  \fmf{plain,label=$\phi$} {i1,v1}
  \fmf{plain,label=$\phi$} {v1,i2}
  \fmf{plain,label=$\phi$} {v1,o1}
  \fmf{plain,label=$\phi$} {o2,v1}
  \fmfright{o1,o2} \fmflabel{$p_4$}{o1} \fmflabel{$p_3$}{o2}
  \fmfdot{v1}  
\end{fmfchar*}
\end{fmffile} \vspace{10pt} $$ \frac{\mathrm i \lambda^2} 2 \sum_{\sigma_i \in \{\vec p_1, \vec p_2, \vec p_3, \vec p_4 \}} \frac{1}{-\mod{\sigma_1 + \sigma_2} + m^2}$$
    \end{center}
  \end{minipage}}
  \subfloat[]{\label{IOE2c}\begin{minipage}{0.3\linewidth}
    \begin{center}
    \unitlength = 1mm
    \begin{fmffile}{IOE23}
  \begin{fmfchar*}(25,25)
\fmfleft{i1}
\fmfright{o1}
\fmf{plain,tension=3}{i1,v1}
\fmf{plain,tension=3}{v1,o1}
\fmf{plain,right,tension=0.6}{v1,v1}
\fmfdot{v1}
\end{fmfchar*}
\end{fmffile}
\vspace{9pt}
    \end{center}
  \end{minipage}}
\caption[justification=justified,singlelinecheck=false]{(a) and (b) The Feynman rules for the reduced Lagrangian (\ref{Lred}); 
(c) One-loop diagrams for the $\phi$ propagator. }
\label{fig:IOE2}
	    \vspace{9pt}
\end{figure}

To be able to neglect the tadpoles,  we need to understand how they manifest themselves when we integrate out the offending fields. To this end we integrate out the field $A$ in the Lagrangian (\ref{intoutexlag}) using the constraint (\ref{conA}). Substituting the constraint back in we obtain
\beq
\L_{reduced} = - \half (\del_\mu \phi)^2 + \frac{\lambda^2} 2 \phi^2 \frac 1 {\triangle + m^2} \phi^2, \label{Lred}
\eeq
where the term containing the inverse of $\triangle$ is formally defined using a Fourier transformation. The resulting equation of motion is given by
\beq
\frac{\delta}{\delta \phi} \int d^4 x \L_{reduced}=\Box \phi+2 \lambda^2 \phi \frac 1 {\triangle + m^2} \phi^2=0
\eeq
Note that one obtains exactly the same equation from substituting the constraint (\ref{conA}) into the $\phi$ equation of motion (\ref{peq}), thereby illustrating the consistency of our method. The Feynman rules for the reduced Lagrangian (\ref{Lred}) are now shown\footnote{Note the permutations of $\sigma_1 \neq \sigma_2$ across the set of $\vec p_i$s.} in Figure \ref{fig:IOE2}, along with the only one-loop contribution to the propagator correction.
Computing our solitary Feynman diagram, we obtain
\beq
\half \left( \frac{- \mathrm i}{k^2} \right)^2 \int \frac{d^4 p}{(2 \pi)^4} \left( \frac{- \mathrm i}{p^2} \right) \frac{\mathrm i \lambda^2} 2 \sum_{\mathrm{perms}} \frac 1 {- \mod{\vec p_3 + \vec p_4}^2 + m^2}.
\eeq
Before proceeding further, we need to consider all the permutations.
Essentially, we need to find all the permutations pairing elements of the set $\{ \vec p_1,\vec p_2,\vec p _3,\vec p_4\}$ with $\{ \vec k, -\vec k, \vec p, - \vec p \}$, the momenta of each leg of the vertex in the relevant diagram.
Eight permutations result in $\vec p_3 + \vec p_4 = \vec p + \vec k$, four in $\vec p_3 + \vec p_4 = \vec p - \vec k$, four in $\vec p_3 + \vec p_4 = - \vec p + \vec k$ and eight in $\vec p_3 + \vec p_4 = 0$.
Using the fact we are integrating over $p$ and only care about the modulus squared, we can rewrite these as sixteen giving $\vec k - \vec p$ and eight permutations giving $0$.
So, the one-loop correction to the propagator gives
\beq\label{IOEC}
 \frac{- \lambda^2}{k^4} \int \frac{d^4 p}{(2 \pi)^4} \frac 1 {p^2} \left[ \frac 4 {-\mod{\vec k-\vec p}^2 + m^2} + \frac 2 {m^2} \right].
\eeq
Clearly, the first term in \eqref{IOEC} is equal to the 1PI  contribution \eqref{IOEA} derived earlier, while the second is equal to the tadpole contribution \eqref{IOEB}. Therefore, if we want to neglect the tadpole contributions for the reasons described above we need to take care with ``bubblegum diagrams" with the generic shape shown in Figure \ref{fig:IOE2}. In particular we should not include permutations that lead to vanishing combinations of momenta in the relevant 4-vertex. Upon integrating the non-dynamical field {\it back in} we now understand this as vanishing momenta being transferred to a loop by the propagator for the non-dynamical field in the tadpole diagram.  We keep this in mind when computing bubblegum diagrams in Ho\v rava gravity.

\subsection{Reduced action for Ho\v rava gravity coupled to a relativistic scalar field}
Our goal is to identify one-loop corrections to the relativistic propagator for a scalar field coupled to Ho\v rava gravity. At tree level, this theory is described by the following action
\beq S=S_{grav}+S_\varphi^{tree}
\eeq
where $S_{grav}$ is given by the action for Ho\v rava gravity (\ref{Sgrav}) and $S_\varphi^{tree}$ is given by the relativistic action (\ref{stree})  for a scalar field of mass $m$ and potential $\varphi^4$, coupled to the spacetime metric.  The Hamiltonian and momentum constraints for this theory are
\begin{subequations}\label{horadmcst}
\begin{align}
{\cal C}_N=\funcd{S}{N} &= M_{pl}^2 \sqrt \gamma \left[ - K_{ij} K^{ij} + \lambda K^2 + R - \alpha a^i a_i - 2 \alpha D_i a^i \right] \nonumber \\
&+ \sqrt \gamma \left[ A_1 R_{ij}^2 + A_2 R^2 + A_3 \left( R D_i a^i + \frac 1 N \triangle (N R) \right) + A_4 \left( (D_i a^i)^2 + \frac 2 N \triangle (N D_i a^i) \right) \right] \nonumber \\
& + \frac{ \sqrt \gamma}{M_{pl}^2} \left[ B_1 (D_i R_{jk})^2 + B_2 (D_i R)^2 + B_3 \left( \triangle R D_i a^i + \frac 1 N \triangle (N \triangle R) \right)
 \right. \nonumber \\ 
 & \qquad \qquad \qquad \left. 
 + B_4 \left( D_i a^i \triangle D_j a^j + \frac 1 N \triangle (N \triangle D_i a^i) + \frac 1 N \triangle^2 (N D_i a^i) \right) \right] \nonumber \\
 & - \sqrt \gamma \left[ \frac 1 {2N^2} \left( \dot \varphi - N^i \del_i \varphi \right)^2 + \frac 1 2 D^i \varphi D_i \varphi +\frac{1}{2} m^2 \varphi^2 +\frac{\mu}{4!} \varphi^4\ \right]  \\
{\cal C}_i=\funcd{S}{N_i} &= 2 M_{pl}^2 \sqrt \gamma \left( D_j K^{ij} - \lambda D^i K \right)+\frac {\sqrt \gamma} N \left[ - \dot \varphi D^i \varphi + D^i \varphi N_j D^j \varphi \right] 
\end{align}
\end{subequations}
We need to establish the form of the reduced action for the dynamical fields, having integrated out the constraints up to the appropriate order. To this end, we begin by perturbing our ADM fields about Minkowski
\begin{subequations}\label{ADMperts}
\begin{align}
N &= 1 + \epsilon n \\
N_i &= \epsilon \left( \del_i \beta + S_i \right) \qquad \text{where } \del^i S_i = 0 \\
\gamma_{ij} &=\delta _{ij} \left( 1 + 2 \frac{\epsilon}{M_{pl}} \zeta \right) + 2 \epsilon \del_i \del_j E + 2 \epsilon \del_{(i} V_{j)} + \frac \epsilon {M_{pl}} h_{ij} \qquad \text{where } \del_i V^i = \del_i h^{ij} = h^i_i = 0,
\end{align}
\end{subequations}
where we have introduced the expansion parameter $\epsilon$, and we have assumed units in which the emergent speed of light $c=1$.
Note that once this expansion has been made, we will not be concerned with distinguishing between upper or lower indices, since they will all be spatial and flat. For the matter sector, we have to ensure we replace $\varphi \to \epsilon \varphi$, since
we are also considering these as leading order perturbations to a vacuum Minkowski background.

Having performed a helicity decomposition on the metric components it is convenient to introduce projection operators, 
\beq
\pi_{ij} \equiv \delta_{ij} - \frac{\del_i \del_j}{\triangle} \qquad
\half \Pi_{ij|kl} \equiv \half \left( 2 \pi_{i(k} \pi_{l)j} - \pi_{ij} \pi_{kl} \right),
\eeq
which project out the transverse and transverse-traceless components respectively.  When we switch to Fourier space, these will have a vector as a superscript, \eg~$\pi^{\vec k}_{ij}$ in which case one replaces $\partial_i \to  \vec k_i$ in the above expressions. We will also find it useful to define
\beq
f(\vec k) := \frac{1 - \frac{A_3}{M_{pl}^2} \vec k^2 + \frac{B_3}{M_{pl}^4} \vec k^4}{\alpha +  \frac{A_4}{M_{pl}^2} \vec k^2 + \frac{B_4}{M_{pl}^4} \vec k^4}
\eeq

Some of the unphysical metric degrees of freedom can be removed by gauge fixing, others we will have to integrate out\footnote{Recall in EM, one can obtain an action in solely the two degrees of freedom by removing the longitudinal part with the transverse gauge fixing $\del^i A_i = 0$ and integrating out the non-dynamical field $A_0$.}. It is clear from the transformations \eqref{fpdifffields} that we may choose the gauge
\beq\label{gfsc}
V_i = 0 \quad E = 0.
\eeq
without losing knowledge of our constraints. We have now reduced our expansion of $\gamma_{ij}$ to the physical scalar and tensor.
The pieces arising from $N$ and $N_i$ will be removed only when we integrate out the corresponding constraints.
Expanding the action order by order in $\epsilon$, we find that
\beq \label{fullscalaraction}
S = \epsilon^2 S^{(2)} + \epsilon^3 S^{(3)} + \epsilon^4 S^{(4)} + \ordep{5}
\eeq
where
\begin{subequations} \label{expquoteactionsc}
\begin{align}
 \label{quadraticquoteactionsc}
S^{(2)} = \int d t d^3 \vec x & \left[ \half \varphi \left( - \del_t^2 + \triangle - m^2 \right) \varphi + \frac 1 4 h_{ij} \left( - \del_t ^2 + \triangle + \frac{A_1}{M_{pl}^2} \triangle^2  - \frac{B_1}{M_{pl}^4} \triangle^3 \right) h_{ij} \right. \nonumber \\
& \left. + M_{pl}^2 n  \left( \alpha \triangle - \frac{A_4}{M_{pl}^2} \triangle^2 + \frac{B_4}{M_{pl}^4} \triangle^3 \right) n
- M_{pl}^2 (1 - \lambda) \beta \triangle^2 \beta + \half M_{pl}^2 S_i \triangle S_i \right.\nonumber  \\
& \left.  + 3 (1 - 3 \lambda) \dot \zeta^2 - 2 \zeta \triangle \zeta + \frac{\left( 6 A_1 + 16 A_2 \right)}{M_{pl}^2} \zeta \triangle^2 \zeta - \frac{\left( 6 B_1 + 16 B_2 \right)}{M_{pl}^2} \zeta \triangle^3 \zeta  +n C_N^{(1)}+n_i C_i^{(1)}\right]
 \\
\label{cubicquoteactionsc} 
S^{(3)} =\frac 1 {M_{pl}} \int d t d^3 \vec x & \Bigg\{ \frac 3 2 \zeta \dot \varphi^2 - \half \zeta \del_i \varphi \del^i \varphi - \frac 3 2 m^2 \zeta \varphi^2 + \half h^{ij} \del_i \varphi \del_j \varphi \nonumber \\
 & + \alpha M_{pl}^2 \left[ - \zeta \del_i n \del_i n + 2 M_{pl} n \del_i n \del_i n \right]
 - 4 A_3 \triangle \zeta \del_i n \del_i n - 4 \frac {B_3} {M_{pl}^2} \triangle^2 \zeta \del_i n \del_i n \nonumber \\
 & + A_4 \left[ \zeta (\triangle n)^2 + 2 M_{pl} n (\triangle n)^2 - 2 \del_i \zeta \del_i n \triangle n + 4 M_{pl} \del_i n \del_i n \triangle n \right] \nonumber \\
 & + \frac{B_4}{M_{pl}^2} \left[ \zeta \triangle n \triangle^2 n - \del_i \zeta \del_i n \triangle^2 n + 2 M_{pl} \del_i \del_i n \triangle^2 n - \del_i \zeta \del_i \triangle n \triangle n \right. \\ & \qquad \qquad \left. + 2 \triangle n \triangle ( \zeta \triangle n) + 2 M_{pl} \triangle n \triangle (n \triangle n) - \triangle n \triangle(\del_i \zeta \del_i n) + 2 M_{pl} \triangle n \triangle(\del_i n \del_i n)  \right]\nonumber  \\
 & + 2 M_{pl}^3 n \left( \del_i \del_j \beta \right)^2 - 2 M_{pl}^3 \lambda n (\triangle \beta)^2 - 2 M_{pl}^2 (1-3 \lambda) \dot \zeta n \triangle \beta  + 2 M_{pl}^3 n \del_{(i} S_{j)} \del_i S_j \nonumber \\
& + 4  M_{pl}^3 n \del_i \del_j \beta \del_i S_j +  M_{pl}^2 \zeta (\del_i \del_j \beta + \del_{(i} S_{j)}) (\del_i \del_j \beta + \del_{(i} S_{j)}) -  M_{pl}^2 \lambda \zeta (\triangle \beta)^2  \nonumber \\
& + 4  M_{pl}^2 \del_i \zeta (\del_j \beta + S_j ) (\del_i \del_j \beta + \del_{(i} S_{j)} ) - 2  M_{pl}^2 (1 - \lambda) \del_i \zeta (\del_i \beta + S_i) \triangle \beta +M_{pl} \left(n C_N^{(2)}+n_i C_i^{(2)}\right)  \Bigg \} \nonumber \\
&+\ldots, \nonumber \\
 \label{quarticquoteactionsc}
S^{(4)} =\frac 1 {M_{pl}^2} \int \d t d^3 \vec x & \Bigg\{ \frac 3 4 \zeta^2 \dot \varphi^2 + \frac 1 4 \zeta^2 \del_i \varphi \del^i \varphi - \frac 3 4 m^2 \zeta^2 \varphi^2 + \frac 3 2 \zeta h^{ij} \del_i \varphi \del_j \varphi \nonumber \\
& - \frac 1 8 h_{ij} h^{ij} \dot \varphi^2 + \frac 1 8 h_{ij} h^{ij} \del_k \varphi \del^k \varphi - \half h^{ik} h^{j}_k \del_i \varphi \del_j \varphi + \frac 1 8 m^2 h^{ij} h_{ij} \varphi^2 \\
& - \half M_{pl}^2 n^2 \dot \varphi^2 -  M_{pl}^2 n n^i \dot \varphi \del_i \varphi - \half  M_{pl}^2 n^i n^j \del_i \varphi \del_j \varphi - \frac 1 {4!} M_{pl}^2 \mu \varphi^4 +M_{pl}^2\left(n C_N^{(3)}+n_i C_i^{(3)}\right)\Bigg  \} \nonumber  \\
&+\ldots, \nonumber 
\end{align}
\end{subequations}
and ``$\ldots$" denote terms which are irrelevant to our subsequent calculations, and the constraints are expanded as $\mathcal C_N= \epsilon C_N^{(1)} +\epsilon^2 C_N^{(2)}+\epsilon^3 C_N^{(3)}+ \ordep{4}$ and  $\mathcal C_i= \epsilon C_i^{(1)} +\epsilon^2 C_i^{(2)}+\epsilon^3 C_i^{(3)}+ \ordep{4}$.  Note that we do not need to consider interactions beyond fourth order since  for 1-loop corrections to the propagator we will only encounter up to four point vertices. 

We now integrate out the constraints by setting $C_N^{(1)}=-\epsilon C_N^{(2)}-\epsilon^2 C_N^{(3)}-\epsilon^3 C_N^{(4)}+\ordep{4}$ and $C_i^{(1)}=-\epsilon C_i^{(2)}-\epsilon^2 C_i^{(3)}-\epsilon^3 C_i^{(4)}+\ordep{4}$, or more specifically,

\begin{subequations}\label{scalarplusgravthreeconstraints}
\begin{align}
2 \epsilon M_{pl}^2 \left( - \alpha + \frac{A_4 \triangle}{M_{pl}^2} - \frac{B_4 \triangle^2}{M_{pl}^4} \right) \triangle n &= 4 \epsilon M_{pl} \left(1 + \frac{A_3 \triangle}{M_{pl}^2} + \frac{B_3 \triangle^2}{M_{pl}^4} \right) \triangle \zeta \nonumber \\
& \qquad + \frac {\epsilon^2} {2} \left(\dot \varphi^2 + \del_i \varphi \del^i \varphi + m^2 \varphi^2 \right)  + \epsilon^2 H_2  + \epsilon^3 H_3  \label{scalarplusgravthreeconstraintsn} \\
& \qquad - \epsilon^3 \left[ n \dot \varphi^2 + n^i \del_i \varphi \dot \varphi + \half h^{ij} \del_i \varphi \del_j \varphi + \zeta \del_i \varphi \del^i \varphi \right] + \ordep 4  \nonumber \\
2 \epsilon M_{pl}^2 (1 - \lambda) \triangle^2 \beta  &= 2 \epsilon M_{pl} (1 - 3 \lambda) \triangle \dot \zeta - \epsilon^2 \del^i \left( \dot \varphi \del_i  \varphi \right) + \epsilon^2 P_2 \nonumber \\
& \qquad + \epsilon^3 \del^i \left( \del_i \varphi \del_j \varphi (\del^j \beta + S^j) \right) + \epsilon^3 P_3 + \ordep 4  \label{scalarplusgravthreeconstraintsbeta} \\
\epsilon M_{pl}^2 \triangle S_i &= - \epsilon^2 \pi_{ij} \dot \varphi \del^j \varphi + \epsilon^2 {Q_2}_i \nonumber \\ & \qquad + \epsilon^3 \pi_{ij} \left( \del^j \varphi \del^k \varphi (\del_k \beta + S_k ) \right)  + \epsilon^3 {Q_3}_i + \ordep 4  \label{scalarplusgravthreeconstraintss}
\end{align}
\end{subequations}
where 
\begin{subequations}
\begin{align}
H_q &= H_q \left( h_{ij}, \zeta, n, \beta, S_i \right) = - \deriv{^q}{\epsilon^q} \left( \frac 1 {q!} \funcd{S_{grav}}{N} \right) \bigg|_{\epsilon=0} \label{Hq} \\
P_q &= P_q \left( h_{ij}, \zeta, n, \beta, S_i \right) = \del^i \deriv{^q}{\epsilon^q} \left( \frac {N \gamma_{ij}} {q!} \funcd{S_{grav}}{N_j} \right) \bigg|_{\epsilon=0} \label{Pq}  \\
Q_q^i &= Q_q^i \left( h_{ij}, \zeta, n, \beta, S_i \right) = \pi^{ij} \deriv{^q}{\epsilon^q} \left( \frac {N \gamma_{jk}} {q!} \funcd{S_{grav}}{N_k} \right) \bigg|_{\epsilon=0}. \label{Qq}
\end{align}
\end{subequations}
Of course, we obtain three equations from the two constraints as the momentum constraint can be split into its transverse and longitudinal parts, yielding two equations. Note that  $H_q, P_q$ and $Q_q$  contain $n$, $\beta$ and $S_i$, which can be removed iteratively by re-substituting \eqref{scalarplusgravthreeconstraints} into the resulting expression.  

We now use equations \eqref{scalarplusgravthreeconstraints} to eliminate the non-dynamical field $n, \beta$ and $S_i$ from the action, thereby arriving at the reduced action for the dynamical fields, $h_{ij}, \zeta$ and $\varphi$. 
\begin{multline}
S_{reduced}=\int dt d^3 \vec x~ \epsilon^2 \left[\frac{1}{2} h_{ij} O^{ij |kl} h_{kl} +\frac{1}{2} \zeta O^\zeta \zeta+\frac{1}{2} \varphi(-\del_t^2+\triangle-m^2)\varphi\right]\\+\epsilon^3\left[V_{h \varphi^2}+V_{\zeta\varphi^2}+\ldots \right] +\epsilon^4 \left[ V_{\varphi^4}+
V_{h^2 
\varphi^2} +V_{\zeta^2 
\varphi^2}+\ldots \right]+\ordep{5}
\end{multline}
where $O^{ij|kl}$ and $O^\zeta$ denote complicated operators for the leading order kinetic terms for $h_{ij}$ and $\zeta$. There are two important three point vertices and three important four point vertices: the $h_{ij} \varphi^2$ vertex denoted by $V_{h \varphi^2}$; the $\zeta \varphi^2$ vertex denoted by $V_{\zeta \varphi^2}$; the $\varphi^4$ vertex denoted by $V_{\varphi^4}$; the  $h_{ij} h_{kl} \varphi^2$ vertex denoted by $V_{h^2  \varphi^2}$; and the $\zeta^2 \varphi^2$ vertex denoted by $V_{\zeta^2 \varphi^2}$. Again, the ``$\ldots$" correspond to terms that will play no role in  the 1-loop correction to the scalar propagator, namely pure gravity vertices.

\subsubsection{Feynman Rules}
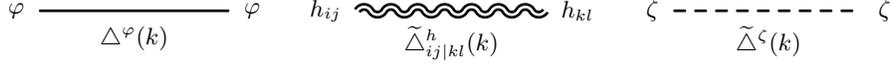
\begin{figure}[t]
	    \vspace{9pt}
  \begin{minipage}{1.00\linewidth}
    \begin{center}
    \unitlength = 1mm
\begin{fmffile}{particledef}
\begin{fmfchar*}(25,25)
  \fmfleft{i1}  \fmflabel{$\varphi$}{i1}
  \fmf{plain,label=$\triangle^\varphi(k)$} {i1,o1}
  \fmfright{o1}  \fmflabel{$\varphi$}{o1}
\end{fmfchar*}
\hspace{1.5cm}
\begin{fmfchar*}(25,25)
  \fmfleft{i1}\fmflabel{$h_{ij}$}{i1}
  \fmf{dbl_wiggly,label=$\tilde \triangle^h_{ij|kl}(k)$} {i1,o1}
  \fmfright{o1}  \fmflabel{$h_{kl}$}{o1}
\end{fmfchar*}
\hspace{1.5cm}
\begin{fmfchar*}(25,25)
  \fmfleft{i1} \fmflabel{$\zeta$}{i1}
  \fmf{dashes,label=$\tilde \triangle^\zeta(k)$} {i1,o1}
  \fmfright{o1} \fmflabel{$\zeta$}{o1}
\end{fmfchar*}
\end{fmffile}
\caption{Propagators for the dynamical fields}
\label{fig:particledef}
    \end{center}
  \end{minipage}
\end{figure}

\begin{figure}[htb]
	    \vspace{9mm}
  \begin{minipage}{1.00\linewidth}
    \begin{center}
    \unitlength = 1mm
\begin{fmffile}{scalproploopsvert}
\qquad \begin{fmfchar*}(20,20)
  \fmfleft{i1,i2} \fmflabel{$k_1$}{i1} \fmflabel{$k_2$}{i2}
  \fmf{plain} {i1,v1}
  \fmf{plain} {i2,v1}
  \fmf{dbl_wiggly} {o1,v1}
  \fmfright{o1} \fmflabel{$k_3, ij$}{o1}
  \fmfdot{v1} \fmflabel{$V_{ij}(k_1,k_2,k_3)$ }{v1}  \fmfkeep{hp2}
\end{fmfchar*} 
\hspace{4cm}
\begin{fmfchar*}(20,20)
  \fmfleft{i1,i2} \fmflabel{$k_1$}{i1} \fmflabel{$k_2$}{i2}
  \fmf{plain} {i1,v1}
  \fmf{plain} {i2,v1}
  \fmf{dashes} {o1,v1}
  \fmfright{o1} \fmflabel{$k_3$}{o1}
  \fmfdot{v1}  \fmflabel{$V(k_1,k_2,k_3)$ }{v1}
\end{fmfchar*}
\\
\vspace{0.5cm} \hspace{1cm} (a) $V_{h \varphi^2}$ \hspace{4.8cm} (b)  $V_{\zeta \varphi^2}$ 
\\
\vspace{1.5cm}
\begin{fmfchar*}(20,20)
  \fmfleft{i1,i2} \fmflabel{$k_1$}{i1} \fmflabel{$k_2$}{i2}
  \fmf{plain} {i1,v1}
  \fmf{plain} {i2,v1}
  \fmf{plain} {o1,v1}
  \fmf{plain} {o2,v1}
  \fmfright{o1,o2} \fmflabel{$k_3$}{o1} \fmflabel{$k_4$}{o2}
  \fmfdot{v1}  \fmflabel{$W(k_1, k_2, k_3, k_4)$}{v1}
\end{fmfchar*}
\hspace{2.5cm}
\begin{fmfchar*}(20,20)
  \fmfleft{i1,i2} \fmflabel{$k_1$}{i1} \fmflabel{$k_2$}{i2}
  \fmf{plain} {i1,v1}
  \fmf{dbl_wiggly} {o2,v1} 
  \fmf{dbl_wiggly} {o1,v1}
  \fmf{plain} {i2,v1}
  \fmfright{o1,o2} \fmflabel{$k_3, ij$}{o2} \fmflabel{$k_4, kl$}{o1}
  \fmfdot{v1}  \fmflabel{$~V_{ijkl}(k_1,k_2,k_3, k_4)$ }{v1}
\end{fmfchar*}
\hspace{2.5cm}
\begin{fmfchar*}(20,20)
  \fmfleft{i1,i2} \fmflabel{$k_1$}{i1} \fmflabel{$k_2$}{i2}
  \fmf{plain} {i1,v1}
  \fmf{plain} {i2,v1}
  \fmf{dashes} {o1,v1}
  \fmf{dashes} {o2,v1}
  \fmfright{o1,o2} \fmflabel{$k_3$}{o1} \fmflabel{$k_4$}{o2}
  \fmfdot{v1} \fmflabel{$~{\cal V}(k_1,k_2,k_3, k_4)$ }{v1}
\end{fmfchar*}
\end{fmffile} 
\\
\vspace{0.5cm}  (c) $V_{\varphi^4}$ \hspace{3cm} (d)  $V_{h^2\varphi^2}$ \hspace{3cm} (e)  $V_{\zeta^2\varphi^2}$
\\
	    \vspace{9mm}
\caption{Three and four point vertices for the dynamical fields. The precise form of these is presented in appendix \ref{sec:appvert}.}
\label{fig:proploops}
    \end{center}
  \end{minipage}
\end{figure}
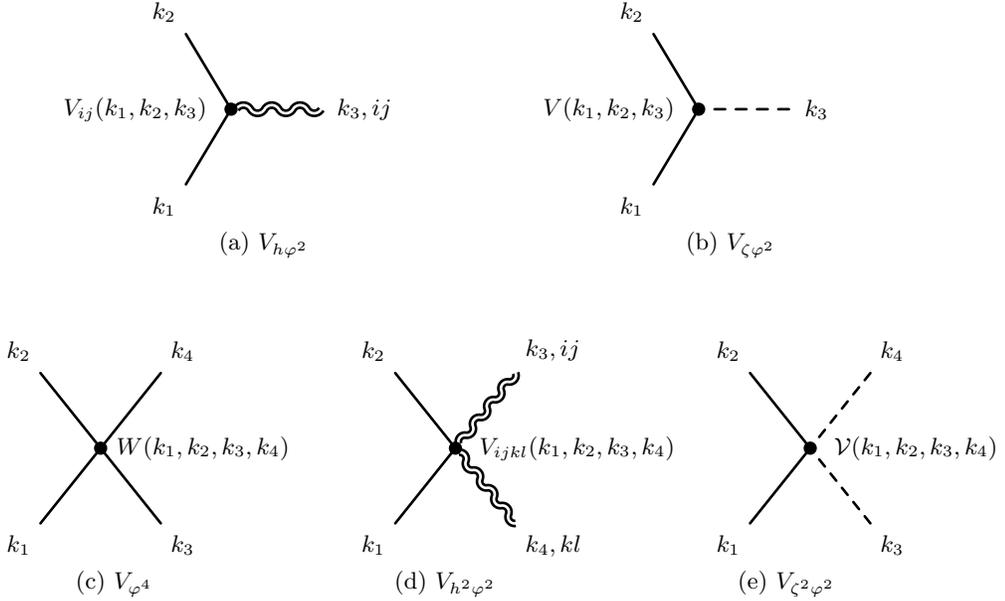

The precise form of these operators is best expressed in terms of the corresponding Feynman rules. Working in Fourier space with four momentum $k^\mu$ split  into energy $\omega_{\vec k}$ and three-momentum $\vec k$, we have the following tree-level propagators, as  shown in Figure \ref{fig:particledef}:
\begin{subequations}
\begin{align}
 i \tilde \triangle^\varphi(k) &= \frac 1 {-\omega_{\vec k}^2 + \abs k^2 + m^2} \\
i \tilde \triangle^h_{ij|kl} (k)&= \frac{\half \Pi^{\vec k}_{ij|kl}}{-\omega_{\vec k}^2 + \abs k^2 + \frac{A_1}{M_{pl}^2} \abs k^4  + \frac{B_1}{M_{pl}^4} \abs k^6} =:  \half \Pi^{\vec k}_{ij|kl} i \tilde \triangle^h (k) \\
-i \left( \tilde \triangle^\zeta (k) \right)^{-1} &= - \frac{3 \lambda - 1}{\lambda - 1} \omega_{\vec k}^2 + \frac 2 {\alpha + \frac{A_4}{M_{pl}^2} \abs k^2 + \frac{B_4}{M_{pl}^4} \abs k^4} \left[ (2 - \alpha) \abs k^2 - \left( A_4 + 4 A_3 \right) \frac{\abs k^4}{M_{pl}^2} \right. \nonumber \\  & \hspace{60mm} \left.   + \left( 4 B_3 - B_4 + 2 A_3 ^2 \right) \frac{\abs k^6}{M_{pl}^4} - 4 A_3 B_3 \frac{\abs k^8}{M_{pl}^6} + 2 B_3^2 \frac{\abs k^{10}}{M_{pl}^8}  \right] \label{zetaprop}  ,
\end{align}
\end{subequations}
where the indexless $\tilde \triangle^h$ has been introduced to allow us to separate the projection operator and the Green's function.
The relevant vertices are shown in Figure \ref{fig:proploops}. Our convention is that all momenta point {\it into} the vertex. The detailed form for each vertex is presented in Appendix \ref{sec:appvert}.

\subsection{One-loop corrections to a scalar field propagator in  Ho\v rava gravity}\label{sec:quanscal-1lcp}
\begin{figure}[htb]
	    \vspace{9mm}
  \begin{minipage}{1.00\linewidth}
    \begin{center}
    \unitlength = 1mm
\begin{fmffile}{scalproploops}
\begin{fmfchar*}(20,20)
\fmfleft{i1}
\fmfright{o1}
\fmflabel{$\quad =$}{o1}
\fmf{plain,tension=3}{i1,v1}
\fmf{plain,tension=3}{v1,o1}
\fmfblob{1cm}{v1}
\end{fmfchar*}
\hspace{1.2cm}
\begin{fmfchar*}(20,20)
\fmfleft{i1}
\fmfright{o1}
\fmflabel{$\quad +$}{o1}
\fmf{plain,tension=3}{i1,v1}
\fmf{plain,tension=3}{v1,o1}
\fmf{plain,right,tension=0.6}{v1,v1}
\fmfdot{v1}
\end{fmfchar*}
\hspace{1.2cm}
\begin{fmfchar*}(20,20)
\fmfleft{i1}
\fmfright{o1}
\fmflabel{$\quad +$}{o1}
\fmf{plain,tension=3}{i1,v1}
\fmf{plain,tension=3}{v2,o1}
\fmf{plain,tension=3}{v1,v2}
\fmf{dbl_wiggly,left,tension=-3}{v1,v2}
\fmf{phantom}{v1,v2}
\fmfdot{v1,v2}
\end{fmfchar*}
\hspace{1.2cm}
\begin{fmfchar*}(20,20)
\fmfleft{i1}
\fmfright{o1}
\fmflabel{$\quad +$}{o1}
\fmf{plain,tension=3}{i1,v1}
\fmf{plain,tension=3}{v1,o1}
\fmf{dbl_wiggly,right,tension=0.6}{v1,v1}
\fmfdot{v1}
\end{fmfchar*}
\hspace{1.2cm}
\begin{fmfchar*}(20,20)
\fmfleft{i1}
\fmfright{o1}
\fmf{plain,tension=3}{i1,v1}
\fmf{plain,tension=3}{v2,o1}
\fmf{plain,tension=3}{v1,v2}
\fmf{dashes,left,tension=-3}{v1,v2}
\fmf{phantom}{v1,v2}
\fmfdot{v1,v2}
\end{fmfchar*}
\hspace{1.2cm}
\begin{fmfchar*}(20,20)
\fmfleft{i1}
\fmfright{o1}
\fmf{plain,tension=3}{i1,v1}
\fmf{plain,tension=3}{v1,o1}
\fmf{dashes,right,tension=0.6}{v1,v1}
\fmflabel{$\quad +$}{i1}
\fmfdot{v1}
\end{fmfchar*}
\end{fmffile}
\label{fig:scalproploops}
    \end{center}
  \end{minipage}
\caption{1-Loop corrections to the scalar propagator}
\label{fig:scalpropcorrs}
\vspace{9mm}
\end{figure}
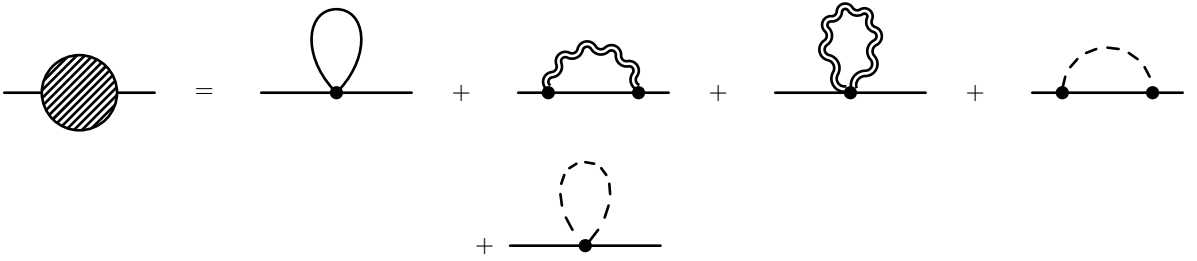

We are now ready to compute the one loop correction to the scalar propagator. To this end, the relevant 1PI graphs are shown in Figure \ref{fig:scalpropcorrs}.
%
As usual, the renormalised two-point vertex for the scalar $\Gamma_{\varphi\varphi}^{ren}=\Gamma^{tree}_{\varphi\varphi}-\Sigma$, where $\Gamma^{tree}_{\varphi\varphi}=(\Delta^\varphi)^{-1}$ is the tree-level vertex and $\Sigma$ is the self energy (at one loop).

Let us now compute the contributions to the self energy for each diagram. Our expressions will be given in terms of the integrations over internal momenta although we will explicitly drop terms that will obviously vanish when this integration is performed --- \eg~terms linear in $\omega_{\vec p}$.

\begin{figure}[htb]
	    \vspace{9pt}
    \unitlength = 1mm
\begin{fmffile}{phi4diag}
\begin{fmfchar*}(25,25)
\fmfleft{i1}
\fmfright{o1}
\fmflabel{$\varphi$}{i1}
\fmflabel{$\varphi$}{o1}
\fmf{plain,tension=3,label=$k$}{i1,v1}
\fmf{plain,tension=3,label=$k$}{v1,o1}
\fmf{plain,right,tension=0.6, label=$p$}{v1,v1}
\fmfdot{v1}
\end{fmfchar*}
\end{fmffile}
\caption{The pure scalar bubblegum diagram with a $\varphi^4$ vertex}
\label{fig:4pointphi}
\end{figure}
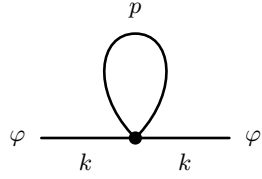

We begin with the pure scalar bubblegum diagram shown in Figure \ref{fig:4pointphi}. 
The appropriate contraction of the legs introduces a symmetry factor of two, so we find that the contribution to the self-energy is given by $\Sigma_{\varphi^4}$, where%
\begin{multline}
\label{phiphiphiphicont}
M_{pl}^2 \Sigma_{\varphi^4}=M_{pl}^2  \int d\omega_{\vec p} d^3 \vec p \tilde \Delta^\varphi (p) \frac{W(k, -k, p, -p)}{ 2} =
     \int d\omega_{\vec p} d^3 \vec p  \tilde \Delta^\varphi (p) \Bigg[ - \frac 1 2  \frac {\omega_{\vec k}^2 \omega_{\vec p}^2 + \left(\vec k \cdot \vec p + m^2 \right)^2} {\alpha \mod{\vec p + \vec k}^2 + \frac{A_4}{M_{pl}^2} \mod{\vec p + \vec k}^4 + \frac{B_4}{M_{pl}^4} \mod{\vec p + \vec k}^6}
      \\ 
    -  \frac {\left( \omega_{\vec p}^2 \vec k^2 + \omega_{\vec k}^2 \vec p^2 \right)} {\mod{\vec k + \vec p}^2} 
    + \left( 1 - \frac 1 {2 (1 - \lambda)} \right) \frac{\left[ (\vec k + \vec p) \cdot \vec k \right]^2 \omega_{\vec p}^2 + \left[ (\vec k + \vec p) \cdot \vec p \right]^2 \omega_{\vec k}^2 }{\mod{\vec k + \vec p}^4} - \frac \mu 2 M_{pl}^2\Bigg],
\end{multline}
Note that since we are computing a bubblegum diagram we have taken care to neglect  the `tadpole-like' contributions as discussed in  section \ref{sec:intoutex}.

 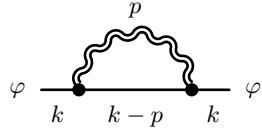
\begin{figure}[htb]
	    \vspace{9pt}
    \unitlength = 1mm
\begin{fmffile}{phi2hX}
\begin{fmfchar*}(25,25)
\fmfleft{i1}
\fmfright{o1}
\fmflabel{$\varphi$}{i1}
\fmflabel{$\varphi$}{o1}
\fmf{plain,tension=3,label=$k$}{i1,v1}
\fmf{plain,tension=3,label=$k$}{v2,o1}
\fmf{plain,tension=3,label=$k-p$}{v1,v2}
\fmf{dbl_wiggly,left,tension=-3,label=$p$}{v1,v2}
\fmf{phantom}{v1,v2}
\fmfdot{v1,v2}
\end{fmfchar*}
\end{fmffile}
\caption{Diagram containing two $h_{ij} \varphi^2$ vertices.}
\label{fig:3pointphih}
\end{figure}
Next  we consider the diagram  containing $h_{ij} \varphi^2$ vertices, shown in  Figure \ref{fig:3pointphih}. As this is not a bubblegum diagram we don't need to worry about tadpole effects. Taking into account the symmetries we find that the contribution to the self energy is $\Sigma_{h\varphi^2}$, where
\begin{multline}\label{phiphihvertexcont}
M_{pl}^2 \Sigma_{h\varphi^2}=M_{pl}^2 \int d\omega_{\vec p} d^3 \vec p V_{ij} (k, p-k, -p) V_{kl}(-k, k-p, p) \tilde \triangle^{\varphi} ( k-p) \tilde \triangle^h_{ijkl} (p) \\= \int d\omega_{\vec p} d^3 \vec p\left[- \half \Pi_{ij|kl}^{\vec p} \vec k_i \vec k_j \vec k_k \vec k_l \tilde \triangle^{\varphi} ( k-p) \tilde \triangle^h (p)\right]
\end{multline}

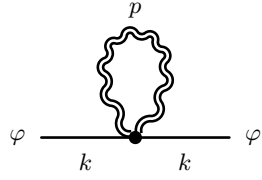
\begin{figure}[htb]
	    \vspace{9pt}
    \unitlength = 1mm
\begin{fmffile}{phi2h2}
\begin{fmfchar*}(25,25)
\fmfleft{i1}
\fmfright{o1}
\fmflabel{$\varphi$}{i1}
\fmflabel{$\varphi$}{o1}
\fmf{plain,tension=3,label=$k$}{i1,v1}
\fmf{plain,tension=3,label=$k$}{v1,o1}
\fmf{dbl_wiggly,right,tension=0.6, label=$p$}{v1,v1}
\fmfdot{v1}
\end{fmfchar*}
\end{fmffile}
\caption{Bubblegum diagram with a tensor graviton in the loop and a  $h_{ij} h_{kl} \varphi^2$ vertex.}
\label{fig:4pointphih}
\end{figure}
Now we consider another bubblegum diagram, this time with the tensor graviton propagating around the loop, as shown in Figure \ref{fig:4pointphih}. The diagram contains a  $h_{ij} h_{kl} \varphi^2$ vertex and, given the symmetry factor of two, contributes a self-energy $\Sigma_{h^2 \varphi^2}$ where
\begin{multline}\label{phiphihhvertexcont}
M_{pl}^2 \Sigma_{h^2 \varphi^2}=\half M_{pl}^2 \int d\omega_{\vec p} d^3 \vec p V_{ijkl} (k, -k; p, -p) \cdot  \tilde \triangle^h_{ijkl} (p) =  \int d\omega_{\vec p} d^3 \vec p \tilde \triangle^h (p) \left[ \frac 1 2 \left( - \omega_{\vec k}^2 + \abs k^2 + m^2 \right) - \pi^{\vec p}_{ij} \vec k_i \vec k_j \right]
\end{multline}

\begin{figure}[htb]
	    \vspace{9pt}
    \unitlength = 1mm
\begin{fmffile}{phi2zetaX}
\begin{fmfchar*}(25,25)
\fmfleft{i1}
\fmfright{o1}
\fmflabel{$\varphi$}{i1}
\fmflabel{$\varphi$}{o1}
\fmf{plain,tension=3,label=$k$}{i1,v1}
\fmf{plain,tension=3,label=$k$}{v2,o1}
\fmf{plain,tension=3,label=$k-p$}{v1,v2}
\fmf{dashes,left,tension=-3,label=$p$}{v1,v2}
\fmf{phantom}{v1,v2}
\fmfdot{v1,v2}
\end{fmfchar*}
\end{fmffile}
\caption{Diagram containing two $\zeta \varphi^2$ vertices}
\label{fig:3pointphizeta}
\end{figure}
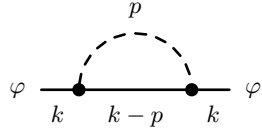
The diagram with the $\zeta \varphi^2$ vertices is shown  in Figure \ref{fig:3pointphizeta}. With the appropriate symmetry factors this gives a self-energy contribution  $ \Sigma_{\zeta \varphi^2}$ where
\begin{eqnarray}
 M_{pl}^2 \Sigma_{\zeta \varphi^2}&=& M_{pl}^2 \int d\omega_{\vec p}d^3 \vec p  V(k, p-k, -p) V(-k, k-p, p) \tilde \triangle^{\varphi} ( k-p) \tilde \triangle^\zeta (p) \nonumber  \\ 
&=&  \int d\omega_{\vec p}d^3 \vec p \tilde \triangle^{\varphi} (k-p) \tilde \triangle^\zeta (p)    
\Bigg[
 \left(3 + 2 f(\vec p)
      \right) \omega_{\vec k} \left( \omega_{\vec k} - \omega_{\vec p}
      \right) - \left(1 - 2 f(\vec p) \right) \vec k \cdot \left( \vec k
      - \vec p \right)  \nonumber
 \\ && \qquad - \left( 3 - 2 f(\vec p) \right) m^2  + \frac{1
      - 3 \lambda}{1 - \lambda} \left[ \frac{\omega_{\vec p}^2}{\abs
      p^2} \dotp p k + \frac{\omega_{\vec p} \omega_{\vec k}}{\abs p^2}
      \left(  \abs p^2 -2 \dotp p k \right) \right]
 \Bigg]^2 \label{phiphizetavertexcont}
\end{eqnarray} 

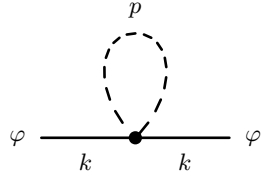
\begin{figure}[htb]
	    \vspace{9pt}
    \unitlength = 1mm
\begin{fmffile}{phi2zeta2}
\begin{fmfchar*}(25,25)
\fmfleft{i1}
\fmfright{o1}
\fmflabel{$\varphi$}{i1}
\fmflabel{$\varphi$}{o1}
\fmf{plain,tension=3,label=$k$}{i1,v1}
\fmf{plain,tension=3,label=$k$}{v1,o1}
\fmf{dashes,right,tension=0.6, label=$p$}{v1,v1}
\fmfdot{v1}
\end{fmfchar*}
\end{fmffile}
\caption{Bubblegum diagram with a scalar graviton in the loop and a $\zeta^2 \varphi^2$ vertex}
\label{fig:4pointphizeta}
\end{figure}
Finally, we consider a third bubblegum diagram, shown in Figure \ref{fig:4pointphizeta}. This has the scalar graviton running through the loop with a  $\zeta^2 \varphi^2$ vertex. Taking care to neglect ``tadpole-like" contributions, we find that the contribution to the self-energy is given by $\Sigma_{\zeta^2 \varphi^2}$ where 
\begin{eqnarray}\label{phiphizetazetavertexcont}
M_{pl}^2 \Sigma_{\zeta^2 \varphi^2} &=&   M_{pl}^2  \int d \omega_{\vec p} d^3 \vec p  \Delta^\zeta (p) \mathcal V(k,-k;p,-p) \nonumber \\&=&
       \int d \omega_{\vec p} d^3 \vec p  \Delta^\zeta (p) \Bigg[ \half \left( 3+8 f (\vec p)^2 \right) \omega_{\vec k}^2  + \half \left[  1 - 8 f(\vec p) \right] \abs k^2 - 2\left( \frac{1-3\lambda}{1 - \lambda} \right)^2 \omega_{\vec p}^2 \frac{\left( \dotp k p \right)^2}{\abs p^4} + \frac 3 2 m^2 \Bigg]
\end{eqnarray}

We cannot hope to solve these integrals exactly, but we can get a handle on their schematic properties by making some approximations.  We will examine the leading order behaviour at low spatial momentum $k \lesssim M_*$ and  assume for simplicity that the scalar potential vanishes ($m=\mu=0$) and that $|\alpha| \sim |1-\lambda| \ll 1$.  In each case we  Wick rotate to Euclidean signature, and perform the integration  over $w_{\vec p}$ followed by the integration over $\vec p$. For the latter, we approximate $|\vec k \pm \vec p|^2 \approx |\vec k|^2+|\vec p|^2$, so that we can integrate out the angular components. We also split the integration over $|\vec p|$ into different regimes, approximating the integrand accordingly. This will hopefully be evident from the example we will work through shortly.  Before doing so, however, let us quote some useful integral formulae, in particular \cite{thooft,Ramond:1981pw}
\beq
I_n=\int_0^\infty  dz \frac{z^{n}}{z^2+A^2}= \frac{A^{n-1}}{2}  \Gamma\left(\frac{1+n}{2}\right) \Gamma\left(\frac{1-n}{2}\right) 
\eeq
For even integer values of $n=2N$,  this integral gives
\beq \label{I2N}
I_{2N}=\frac{(-1)^NA^{2N-1}\pi}{2}
\eeq
 whereas for odd integer values $n=2N+1$ it is divergent. We can regulate the divergence using dimensional regularization, such that
\beq \label{2N+1}
I_{2N+1}=\lim_{\epsilon\to 0}  \int_0^\infty  \frac{d^{1+\epsilon}z}{\mu^\epsilon} \frac{z^{2N+1}}{z^2+A^2}=  (-1)^N A^{2N} \left[-\frac{2} {\epsilon}+\ln\left(\frac{\mu^2}{4\pi A^2}\right) -\gamma \right]
\eeq
where $\gamma$ is the Euler-Masheroni constant and $\mu$ is the renormalisation scale. We will also make use of the following integral which is finite for integer values of $N$
\beq
 \int_0^\infty dz \frac{z^{2N}}{(z^2+A^2)(z^2+B^2)}=\frac{(-1)^N\pi}{2} \left[\frac{A^{2N-1}-B^{2N-1}}{A^2-B^2}\right]
\eeq

Let us now work through the simplest example to illustrate our methods. Consider $\Sigma_{h^2 \varphi^2}$ as given by the integral expression (\ref{phiphihhvertexcont}). Schematically, we write this as
\beq 
\Sigma_{h^2 \varphi^2} \approx \frac{1}{M_{pl}^2}\left[ \# w_{\vec k}^2  \int d  \bar w_{\vec p} d^3 \vec p \frac{1}{ \bar w_{\vec p}^2+|\vec p|^2X(|\vec p|)} +\# |\vec k|^2 \int d  \bar w_{\vec p} d^3 \vec p \frac{1}{\bar w_{\vec p}^2+|\vec p|^2X(|\vec p|)}\right]
\eeq
where $\#$ denotes (not necessarily equal) numbers of order one, and $X(z)=\#+\#\frac{z^2}{M_{pl}^2} +\#\frac{z^4}{M_h^4}$ with $M_h \sim M_{pl}\alpha^{1/4}$ being the scale of Lorentz violation in the tensor sector \cite{blas2}. Here we are obviously being sloppy with tensor structure and  have used the fact that, upon Wick rotating the energy, $w_{\vec p} \to -i \bar w_{\vec p}$, we have  $\tilde  \Delta^h (p) =\frac{\#}{\bar w_{\vec p}^2+|\vec p|^2X(|\vec p|)}$.  We begin by using equation (\ref{I2N}) to do the integration over $w_\vec p$, and then do the angular integration yielding 
\beq
\Sigma_{h^2 \varphi^2} \approx \frac{1}{M_{pl}^2}\left( \# w_{\vec k}^2 \int_0^\infty d|\vec p| \frac{|\vec p|}{\sqrt{X(|\vec p|)} } +\# |\vec k|^2\int_0^\infty d|\vec p| \frac{|\vec p|}{\sqrt{X(|\vec p|)} }\right)
\eeq 
Now for $|\vec p| \ll M_h$, we have $X \sim \#$, whereas for $|\vec p| \gg M_h$ we have $X \sim \# |\vec p|^4/M_h^4$. Thus we split this integral up into two domains and approximate it as follows
\beq
\int_0^\infty d|\vec p| \frac{|\vec p|}{\sqrt{X(|\vec p|)} }\approx \#  \int_0^{M_h}  d|\vec p| |\vec p| +\# \int_{M_h}^\infty d|\vec p| \frac{M_h^2}{|\vec p| }
\eeq
Note that $ \int_{M_h}^\infty d|\vec p| \frac{M_h^2}{|\vec p| } \approx M_h^2 \int_{0}^\infty  d|\vec p| \frac{|\vec p|}{|\vec p|^2 +M_h^2}- \int_0^{M_h}  d|\vec p| |\vec p| $, and so using the formula (\ref{2N+1}) for $N=0$, we obtain
\beq
\Sigma_{h^2 \varphi^2} \approx \frac{M_h^2 }{M_{pl}^2}\left[ \left(\frac{\#}{\epsilon}+\#\ln\frac{\mu^2}{M_h^2} +\#\right) w_{\vec k}^2 +\left(\frac{\#}{\epsilon}+\#\ln\frac{\mu^2}{M_h^2} +\#\right)  |\vec k|^2\right]  
\eeq
This reveals a logarithmic divergence and finite pieces that simply renormalise the constant part of the light cone, but by an amount that is suppressed by a factor of $\sqrt{\alpha}=\frac{M_h^2}{M_{pl}^2}$. 

Using similar techniques, we arrive at the following approximations for the other contributions to the self energy 
\bea
\Sigma_{\varphi^4} &\approx&  \left(\frac{\#}{\epsilon}+\#\ln\frac{\mu^2}{M_*^2} +\#+\# \frac{|\vec k|^2 }{M_*^2 }\right) w_{\vec k}^2 + \left(\frac{\#}{\epsilon}+\#\ln\frac{\mu^2}{M_*^2} + \#+\# \frac{|\vec k|^2 }{M_*^2 }\right) |\vec k|^2 \\
\Sigma_{h\varphi^2} &\approx&  \left[\#+\# \ln\frac{ |\vec k|^2 }{M_h^2} \right]\frac{ |\vec k|^4}{M_{pl}^2} \\
\Sigma_{\zeta\varphi^2} &\approx&  \frac{1 }{\alpha}\left[ \left(\frac{\#}{\epsilon}+\#\ln\frac{\mu^2}{M_{pl}^2} +\#\right) w_{\vec k}^2 +\left(\frac{\#}{\epsilon}+\#\ln\frac{\mu^2}{M_{pl}^2} +\#\right)  |\vec k|^2\right]  \nonumber \\
&&\qquad +\frac{1}{M_*^2  }\left[\left(\#+\# \ln \frac{|\vec k|^2 }{M_*^2 }\right) w_{\vec k}^4 + \left(\#+\#\ln  \frac{|\vec k|^2 }{M_*^2 }\right)w_{\vec k}^2 |\vec k|^2+\left(\#+\# \ln \frac{|\vec k|^2 }{M_*^2 }\right) |\vec k|^4 \right] \label{contribZp2} \\
\Sigma_{\zeta^2 \varphi^2} &\approx& \alpha \left(\frac{\#}{\epsilon}+\#\ln\frac{\mu^2}{M_h^2} +\frac{\#}{\alpha} \right) w_{\vec k}^2 +\left( 1+\#\alpha \right)\left(\frac{\#}{\epsilon}+\#\ln\frac{\mu^2}{M_h^2} +\#\right)  |\vec k|^2
\eea
where we have set $w_{\vec k}=0$ in the denominator of the integrands for $\Sigma_{h\varphi^2}$ and $\Sigma_{\zeta \varphi^2}$, corresponding to  equations (\ref{phiphihvertexcont}) and (\ref{phiphizetavertexcont}) respectively.

 The first thing to note is that we have at most logarithmic divergences on account of the fact that we have used dimensional regularization.  Focussing on the finite terms it is clear that we generate terms of the form
 \beq
 \frac{1}{\alpha} \varphi \left[a_0 +a_1\frac{\Delta}{M_{pl}^2 }+a_2\frac{\del_t^2 }{M_{pl}^2 }++\ldots\right] \ddot \varphi, \qquad \frac{1}{\alpha}  \varphi \left[ b_0+b_1\frac{\Delta}{M_{pl}^2}+\ldots\right] \Delta \varphi 
 \eeq
 where we have neglected the contribution from the $\ln  \frac{|\vec k|^2 }{M_*^2 }$ as they are not expected to be important when we properly take into account infra-red corrections arising from a non-trivial potential (\ie~$m\neq 0, \mu \neq 0$). There are a number of important features to dwell upon. The first is the potentially large leading order correction to the light cone, of order $\delta c^2 \sim (a_0-b_0)/\alpha \gtrsim 10^{7}$.  This large factor is a direct result of the strong coupling between matter and the scalar graviton and suggests an unpalatable amount of fine tuning of the light cone for different particle species. Of course, the effect may be reproduced in exactly equal measure for all particles in which case there is nothing to worry about. It is beyond the scope of this paper to establish whether or not such an optimistic scenario occurs.
 
 Beyond the leading order terms, we have higher derivatives with an additional Planckian suppression. This is the relevant scale  because the scalar graviton propagator, $\tilde \Delta^\zeta$ only feels the $z=3$ scaling at beyond the Planck scale\footnote{From equation (\ref{zetaprop}) we see that the scalar graviton propagator behaves roughly as $\tilde \Delta^\zeta(k) \sim \frac{\alpha} {w_{\vec k}^2-c^2(|\vec k|) |{\vec k}|^2}$ where $$c(|\vec k|) \sim \begin{cases} 1 &|\vec k|< M_* \\ M_*^2/| \vec k|^2 & M_*<|\vec k|<M_h \\ |\vec k|^2/M_{pl}^2  & |\vec k|>M_h \end{cases}$$. \label{footnote}}. Higher spatial derivatives were anticipated in section \ref{sec:nrm}, and may have been expected from the quadratic divergences that appeared  in \cite{Pospelov:2010mp}. Because they were seen to remove these divergences,  it has been suggested \cite{private} that the inclusion of terms such as $(D_i K_{jk})^2$ will help suppress these operators in the UV, beyond the scales $M_*$ and $M_h$.  However, our integrals are evaluated for low momenta $k<M_*$ so we do not probe the very high energy corrections in this paper.
 
In contrast, we did not anticipate  the  terms $\frac{1}{M_{pl}^2 \alpha} \varphi \Delta \ddot \varphi$ and $\frac{1}{M_{pl}^2 \alpha}  \varphi \del_t^4 \varphi$ in section \ref{sec:nrm}, even though they are compatible with the $\DiffFM$ symmetry. This is because we did not endeavour to generalise  terms involving temporal derivatives, consistent with the original formulation of the gravitational action. However, we now see that such terms are generated by loop corrections, and that they alter the temporal part of the propagator in the UV.  This is dangerous and will generically lead to ghosts. Indeed,  the fourth order time derivative can be identified with a new degree of freedom corresponding to an Ostrogradski ghost\cite{ostro}. 
 
Let us consider this fourth order time derivative more closely. It stems from the  $\Sigma_{\zeta \varphi^2}$ contribution to the self-energy, and in particular the piece proportional to $w_{\vec k}^4$, 
\beq
\Sigma_{\zeta \varphi^2} \supset \frac{i}{2 M_{pl}^2} \bar \omega_{\vec k}^4  \int d \bar \omega_{\vec p} d^3 \vec p  \tilde \triangle^\varphi ( k-p)  \tilde \triangle^\zeta (p) \left( 3 + 2 f(\vec p) \right)^2 \label{wk4}
\eeq
where the $\bar \omega$ indicates explicitly that we have performed a Wick rotation $\omega  \to -i \bar \omega$ on all internal and external energies. In our rough evaluation of this integral, we set $\bar w_{\vec k}=0$ inside the scalar part of the loop. One might worry that this eliminates an important correction, so let's see what happens when we leave it in. The {\it Wick rotated} propagators have the approximate form
\beq
\tilde \triangle^\varphi ( k-p)  \sim \frac{1}{(\bar w_{\vec k} -\bar w_{\vec p})^2 +|\vec k-\vec p|^2+m^2}, \qquad \tilde \Delta^\zeta(p) \sim \frac{\alpha} {\bar w_{\vec p}^2+c^2(|\vec p|) |{\vec p}|^2}
\eeq
where $c(|\vec p|)$ is given in footnote \ref{footnote}. Using the Feynman trick  then integrating over $\bar w_{\vec p}$ we obtain, 
\beq
\Sigma_{\zeta \varphi^2} \supset \frac{\pi i}{4 M_{pl}^2} \alpha \bar \omega_{\vec k}^4  \int_0^1 dx\int d^3 \vec p  \frac{ \left( 3 + 2 f(\vec p) \right)^2}{[x(|\vec k-\vec p|^2+m^2)+(1-x)c^2(|\vec p|) |{\vec p}|^2 +x(1-x)\bar w_{\vec k}^2]^{3/2}}
\eeq
Since we are interested in the role of higher order time derivatives, we may as well set the external $3$ momentum to vanish, $\vec k=0$. Now performing the integration over $x$ and then the angles, we obtain,
\beq
\Sigma_{\zeta \varphi^2} \supset 
\frac{2\pi^2  i}{ M_{pl}^2} \alpha\bar \omega_{\vec k}^4 \int_0^\infty d|\vec p|  \frac{ |\vec p | \left( 3 + 2 f(|\vec p|) \right)^2(\sqrt{| \vec p|^2+m^2}+|c(|\vec p|)| |\vec p|  )}{|c(|\vec p|)| \sqrt{|\vec p|^2+m^2} [(\sqrt{| \vec p|^2+m^2}+|c(|\vec p|)| |\vec p|  )^2+\bar w_{\vec k}^2] }
=\frac{2\pi i}{M_{pl}^2}\alpha \bar \omega_{\vec k}^4   \sum_{n=0}^\infty  \frac{ \bar w_{\vec k}^{2n} }{n!} {\cal I}_n
\eeq
where in the last line we have performed a Taylor expansion about $\bar w_{\vec k}^2=0$, with
\beq
{\cal I}_n=\left(-1\right)^n \int_0^\infty d|\vec p|  \frac{ |{\vec p}| \left( 3 + 2 f(|\vec p|) \right)^2}{|c(|\vec p|)| \sqrt{|\vec p|^2+m^2} (\sqrt{| \vec p|^2+m^2}+|c(|\vec p|)| |\vec p|  )^{2n+1}} 
\eeq
Now the crucial point is that, generically, each of the ${\cal I}_n$ is finite so the Taylor expansion is valid in some neighbourhood of $\bar w_{\vec k}^2=0$. This suggests that  the higher order time derivatives are a real phenomena and not some artifact of our rough approximations\footnote{This is basically saying that the expansion  of the integral about $\bar w_{\vec k}^2=0$ does not contain negative powers of $\bar w_{\vec k}^2$ that cancel off the overall factor of $\bar w_{\vec k}^4$.}. We will discuss the pathological implications of these higher order time derivatives and how they may be avoided in more detail in the next section.

 \section{Discussion}\label{sec:concl}
Ho\v rava gravity has attracted much interest in its gravitational sector.
However, the knottier issue of matter in the theory is still relatively new. In this paper we have looked at both classical and quantum effects of Ho\v rava gravity coupled to matter.

Having reviewed pure Ho\v rava gravity in Section \ref{sec:nrg}, we investigated Ho\v rava-like matter theories in Section \ref{sec:nrm}.
We constructed the most general (at quadratic order around a Minkowski background) \DiffFM invariant action of matter coupled to gravity, obeying the usual power-counting renormalisability conditions used in Ho\v rava gravity and assuming the temporal derivatives are as in the relativistic theory.
We constructed these fields both in the usual ADM composition and the St\"uckelberg formalism.
Using this, it was easy to demonstrate that the only way of coupling matter to gravity but not the new mode (in order to evade Lorentz invariance or Equivalence Principle violations) is the standard Lorentz invariant matter action.

Up to this point, we worked classically. However, in Section \ref{sec:quanscal}, we considered the quantum corrections.  In particular we studied one loop corrections to the propagator for a scalar matter field. Our approach differed somewhat from that of  \cite{Pospelov:2010mp} in that we integrated out the constraints and worked directly with the propagating degrees of freedom. We also used dimensional regularization to (roughly) evaluate our loop integrals thereby eliminating the quadratic and quartic divergences that appeared in  \cite{Pospelov:2010mp}.  These divergences now manifest themselves as large momentum dependent corrections.

This analysis has revealed some potentially worrying features. The first is the large renormalisation of the light cone ($\sim 1/\alpha \gtrsim 10^{7}$) at low energies and momentum. This arises because the scalar graviton couples so strongly to the matter sector and was not noticed in \cite{Pospelov:2010mp}  since they only focussed on divergences. Whether or not this means light cones for different particle species must be fine tuned to one part in $10^{7}$ remains to be seen. Work is under way to repeat our analysis for $U(1)$ gauge fields, and preliminary results are expected to be presented in \cite{ianthesis}. What we can say is that the situation can probably be improved by modification of the Ho\v rava action to include terms such as $\left(D_iK_{jk} \right)^2$, provided they are introduced sufficiently far below the Planck scale. Such terms were originally proposed by \cite{Pospelov:2010mp}  to alleviate quadratic divergences in the relative light cones of different species. Here they will act to modify the propagator for the scalar graviton such that it becomes more weakly coupled to matter with increasing momentum.

The second significant feature revealed by our analysis is the generation of higher order temporal derivatives. These are perfectly compatible with the \DiffFM ~symmetry, but are generically associated with Ostrogradski ghosts \cite{ostro}. Higher order time derivatives are also generated in perturbative General Relativity although the corresponding ghosts have Planckian mass and so do not propagate when the effective theory is valid. In contrast, Ho\v rava gravity is touted as a UV complete theory, rather than an effective theory only valid up to some cut-off, so we can always get a ghost to propagate because we can go to arbitrarily high energies.

Can we avoid this problem by modifying the gravitational part of the action?  This seems unlikely since the origin of the  higher order  time derivatives term  can be traced back to the relativistic matter Lagrangian with minimal coupling to gravity. Indeed, consider the standard action
\beq\label{approxmincupscal}
S \sim \int d^4 x \sqrt{-g} g^{\mu \nu} \del_\mu \varphi \del_\nu \varphi,
\eeq
If one expands $g_{\mu \nu} = \eta_{\mu \nu} + \frac 1 {M_{pl}} h_{\mu \nu}$, then one obtains for the $h_{\mu \nu} \varphi^2$ vertex  $V^{\mu \nu} = \frac 1 {M_{pl}} \left[ k \cdot q \eta^{\mu \nu} - 2 k^{(\mu} q^{\nu)} \right]$, where $k, q$ are the energy-momenta of the scalars and $p$ is the energy-momentum of the graviton. 
Working through, one arrives at the contribution to the scalar propagator of
\beq
\sim \frac 1 {M_{pl}^2} \int d^4 p \mathcal V^{\mu \nu \rho \sigma} (k,p) \tilde \triangle^{\varphi} (k-p) \tilde \triangle^{grav}_{\mu \nu \rho \sigma} (p),
\eeq
where $\mathcal V^{\mu \nu \rho \sigma} (k,p) = \left[ k \cdot q \eta^{\mu \nu} - 2 k^{(\mu} q^{\nu)} \right] \left[ k \cdot q \eta^{\rho \sigma} - 2 k^{(\rho} q^{\sigma)} \right]$ and $\tilde \triangle^{grav}_{\mu \nu \rho \sigma} (p)$ is some generalised graviton propagator (which may be a sum of different helicity propagators, \eg~spin-2 tensor and spin-0 scalar gravitons), and $q = k -p$.

If one splits the spacetime indices ($\mu,\nu,\rho,\sigma$) into temporal ($0$) and spatial ($i,j,k,l$) indices, then $\mathcal V^{ijkl}$, $V^{00ij}$ and $\mathcal V^{0000}$ contain $\omega_{\vec k}^4$. This suggests that  fourth order time derivatives will  generically  be generated. We cannot rule out the possibility that the  details of the graviton propagator may be such that the the $\omega^4$ dependence disappears from the integral. Given the discussion at the end of the previous section,  it seems a little optimistic to expect that this could be achieved by a small modification of the gravitational action in Ho\v rava gravity.

Can we avoid the higher time derivatives by modifying the matter action? Naively one might be a little more optimistic for the following reason. Consider the offending  contribution to the self-energy given by equation (\ref{wk4}) but with the Wick rotated scalar propagator given by
\beq
\tilde \triangle^\varphi ( p)  \sim \frac{1}{\bar w_{\vec p}^2 +{\cal Q}^2(|\vec p|)}, 
\eeq
Working through the analysis as at the end of the previous section we find that
\beq
\Sigma_{\zeta \varphi^2} \supset 
\frac{2\pi^2  i}{ M_{pl}^2} \alpha \bar \omega_{\vec k}^4 \int_0^\infty d|\vec p|  \frac{ |\vec p | \left( 3 + 2 f(|\vec p|) \right)^2( |{\cal Q}(|\vec p|)|+|c(|\vec p|)| |\vec p| )}{|c(|\vec p|)| |{\cal Q}(|\vec p|) |[( |{\cal Q}(|\vec p|)|+|c(|\vec p|)| |\vec p|  )^2+\bar w_{\vec k}^2] }
\eeq
If we imagine that both propagators have a pure $z=3$  scaling \ie~${\cal Q} (|\vec p|) \sim {\cal Q}_0|\vec p| ^3$, $c( |\vec p|) |\vec p| \sim c_0|\vec p| ^3$ and take $f(|\vec p|) \sim f_0$, constant then the integral evaluates as $\propto 1/\bar w_{\vec k}^2$,  so that the higher order time derivatives are eliminated. Of course, given that such terms are generated anyway by quantum corrections perhaps it is natural to consider matter Lagrangians that include an explicit $z=3$ scaling in addition to the leading order relativistic piece. However, the leading order relativistic piece will almost certainly spoil the neat cancellation we have just described which relied on exclusively $z=3$ scalings. This question deserves further investigation, not forgetting  the phenomenological implications of introducing Lorentz violating contributions to the classical matter action.

\acknowledgements{
We would like to thank Thomas Sotiriou, Ed Copeland, Kirill Krasnov, Paul Saffin,  and Nemanja Kaloper and Maxim Pospelov for useful discussions.
AP is funded by a Royal Society University
Research Fellowship and IK by an STFC studentship. }

\newpage

\appendix

\section{Vertices}\label{sec:appvert}\setcounter{figure}{2} 
This section contains the explicit form of the  vertices shown in Figure \ref{fig:proploops}. In some cases, the permutations will be done explicitly. In other cases, this will not be done (for clarity). Where it is not performed, a summation $\sum_\pi$ is written explicitly, with the explicit permutations $\pi$ written under the vertex. Recall that all momentum $k_i$ are ingoing. In this section only, $M_{pl}$ has been set equal to one, it is fairly easy to restore these factors by dimensional analysis.

\subsubsection*{Three point $\varphi h$ vertex --- $V_{h\varphi^2}$}

The diagram for the $h_{ij}\varphi^2$ vertex is shown in Figure \ref{fig:proploops}a
where
$
V_{ij} (k_1,k_2;k_3) = - {\vec k_1}_{(i} {\vec k_2}_{j)}
$
\subsubsection*{Three point $\varphi \zeta$ vertex --- $V_{\zeta\varphi^2}$}
The diagram for the $\zeta \varphi^2$ vertex is shown in Figure \ref{fig:proploops}b, where  
\beq
\begin{split}
V (k_1,k_2;k_3) = & - 3 \omega_1 \omega_2 + \dotp{k_1}{k_2} - 3m^2 \\
& - 2 f(\vec k_3) (\omega_1 \omega_2 + \dotp{k_1}{k_2} - m^2) - \frac{1 - 3 \lambda}{1 - \lambda} \frac {\omega_3} {\abs{k_3}^2} (\vec k_1 + \vec k_2) \cdot (\omega_1 \vec k_2 + \omega_2 \vec k_1)
\end{split}
\eeq

\subsubsection*{Four point $\varphi$ vertex --- $V_{\varphi^4}$}
The diagram for the $\varphi^4$ vertex is shown in Figure \ref{fig:proploops}c, where  
\beq
\begin{split}
W(k_1,k_2,k_3,k_4) = & - \frac 1 {16} \sum_{\pi} \frac 1 {\mod{\vec k_3 + \vec k_4}^2} \frac {\left( \omega_1 \omega_2 + \dotp{k_1}{k_2} - m^2 \right) \left( \omega_3 \omega_4 + \dotp{k_3}{k_4} - m^2 \right)} {\alpha + \frac{A_4}{M_pl^2} \abs{\vec k_3 + \vec k_4}^2 + \frac{B_4}{M_pl^4} \abs{\vec k_3 + \vec k_4}^4  } \\
& + \frac 1 4 \frac 1 {1 - \lambda} \sum_{\pi} \frac 1 {\mod{\vec k_3 + \vec k_4}^4} \omega_1 \omega_3 (\vec k_1 + \vec k_2) \cdot \vec k_2 (\vec k_3 + \vec k_4) \cdot \vec k_4 \\
& - \frac 1 2 \sum_{\pi} \frac 1 {\mod{\vec k_3 + \vec k_4}^2} \omega_1 \omega_3 \pi_{ij}^{\vec k_3 + \vec k_4} {\vec k_2}_i {\vec k_4}_j - \mu
\end{split}
\eeq
where $\sum_\pi$ means you sum over all permutations of $\{ 1,2,3,4\}$.

\subsubsection*{Four point $\varphi h$ vertex --- $V_{h^2\varphi^2}$}
The diagram for the $h_{ij} h_{kl} \varphi^2$ vertex is shown in Figure \ref{fig:proploops}d, where
\beq
\begin{split}
V_{ijkl} (k_1,k_2;k_3,k_4) = &\half \delta_{i(k} \delta_{l)j} \left( \omega_1 \omega_2 - \vec k_1 \cdot \vec k_2 + m^2 \right) + \delta_{jl} \left( {\vec k_1}_i {\vec k_2}_k + {\vec k_1}_k {\vec k_2}_i \right) \\
& + \frac 1 {\mod{\vec k_3 + \vec k_4}^2} \left[ \frac 1 \alpha \langle \mathcal H_2 \rangle \left( \omega_1 \omega_2 + \dotp{k_1}{k_2} - m^2 \right) \right. \\
& \qquad \qquad \qquad \left. - \frac i {1- \lambda} ( \vec k_1 + \vec k_2) \cdot (\vec k_2 \omega_1 + \vec k_1 \omega_2) \langle \mathcal P_2 \rangle + 2 (\omega_1 {\vec k_2}_m + \omega_2 {\vec k_1}_m) \langle \mathcal Q_2 \rangle_m \right]
\end{split}
\eeq
and
\begin{subequations}
\begin{align}
\langle \mathcal H_2 \rangle &= \frac 1 4 \delta_{i(k} \delta_{l)j} \left[ - \omega_3 \omega_4 - A_1 \abs{k_3}^2 \abs{k_4}^2 + B_1 \dotp{k_3}{k_4} \abs{k_3}^2 \abs{k_4}^2 \right] \nonumber \\
 & \qquad + \frac 1 2 \sum_{3 \leftrightarrow 4} \delta_{i(k} \delta_{l)j} \left[ \abs{k_3}^2 + \frac 3 4 \dotp{k_3}{k_4} \right] \left[ 1 - A_3 \mod{\vec k_3 + \vec k_4}^2 + B_3 \mod{\vec k_3 + \vec k_4}^4) \right] \\
 & \qquad - \frac 1 4 \sum_{3 \leftrightarrow 4}  \left[ 1 - A_3 \mod{\vec k_3 + \vec k_4}^2 + B_3 \mod{\vec k_3 + \vec k_4}^4) \right] \delta_{jl} {\vec k_3}_k {\vec k_4}_i
 \nonumber \\
\langle \mathcal P_2 \rangle &= \frac 1 2 \sum_{3 \leftrightarrow 4} (\pm i) \omega_4 \left[ \delta_{jl} {\vec k_4}_i {\vec k_3}_k - \half (\vec k_3 + \vec k_4) \cdot \vec k_3 \delta_{i(k} \delta_{l)j} - \lambda \mod{\vec k_3 + \vec k_4}^2 \delta_{i(k} \delta_{l)j} \right]\\
\langle \mathcal Q_2 \rangle_m &= \frac 1 2 \sum_{3 \leftrightarrow 4} \omega_4 \pi_{mn}^{\vec k_3 + \vec k_4} \left[ - {\vec k_4}_i \delta_{jl} \delta_{nk} + \half {\vec k_3}_j \delta_{i(k} \delta_{l)j} \right]
\end{align}
\end{subequations}

\subsubsection*{Four point $\varphi \zeta$ vertex --- $V_{\zeta^2\varphi^2} $}
The  diagram for the $\zeta^2 \varphi^2$ vertex is shown in Figure \ref{fig:proploops}e, where
\footnotesize
\beq
\begin{split}
\mathcal V (k_1,k_2;k_3,k_4) =
 & - 3 \omega_1 \omega_2 - \dotp{k_1}{k_2} - 3 m^2
  + 8 f(\vec k_3) f (\vec k_4) \omega_1 \omega_2
   - \frac 1 2 \left( \frac{1 - 3 \lambda}{1 - \lambda} \right)^2 \frac{\omega_3 \omega_4}{\abs{k_3}^2 \abs{k_4}^2} \sum_\pi (\dotp{k_1}{k_3}) (\dotp{k_2}{k_4}) \\
 & - 2 \frac{1 - 3 \lambda}{1 - \lambda} \left( f(\vec k_3) \frac{\omega_4}{\abs{k_4}^2} \vec k_4 + f(\vec k_4) \frac{\omega_3}{\abs{k_3}^2} \vec k_3 \right) \cdot \left( \omega_1 \vec k_2 + \omega_2 \vec k_1 \right) 
 - 2 \frac{1 - 3 \lambda}{1 - \lambda} \sum_\pi \frac{\omega_2 \omega_4 \dotp{k_3}{k_1}}{\abs{k_3}^2} f(\vec k_4) \\
  & + \frac 1 {4 \alpha} \frac{\omega_1 \omega_2 + \dotp{k_1}{k_2} - m^2}{\mod{\vec k_1 + \vec k_2}^2} \langle H_2 \rangle + 4 \dotp{k_1}{k_2} \left( f(\vec k_3) + f(\vec k_4) \right) - 16f(\vec k_3) f(\vec k_4) \omega_1 \omega_2 \\
    & - \frac i {1 - \lambda} \frac{(\vec k_1 + \vec k_2)}{\mod{\vec k_1 + \vec k_2}^2} \cdot (\vec k_2  \omega_1 + \vec k_1  \omega_2) \langle P_2 \rangle + 2 \frac{\pi_{ij}^{\vec k_1 + \vec k_2}}{\mod{\vec k_1 + \vec k_2}^2} \left( \omega_1 {\vec k_2}_j + \omega_2 {\vec k_1}_j \right) \langle Q_2 \rangle_i \\
 & + \frac 1 {\mod{\vec k_1 + \vec k_2}^2} (\omega_1 \omega_2 + \dotp{k_1}{k_2} - m^2) \left\{ (\vec k_1 + \vec k_2) \cdot \left[\vec k_3 f(\vec k_3) + \vec k_4 f(\vec k_4) \right] \right. \\ & \qquad \qquad \qquad \qquad \qquad \qquad \qquad \left. + f(\vec k_3) f(\vec k_4)\left[ \dotp{k_3}{k_4} + (\vec k_1 + \vec k_2) \cdot (\vec k_3 + \vec k_4) \right] \right\} \\
 & - \frac 4 \alpha \sum_\pi \left\{ \left[ A_3 \abs {k_4} ^2 - B_3 \abs {k_4} ^2 \right] f (\vec k_3) \frac{\vec k_3 \cdot (\vec k_1 + \vec k_2)}{\mod{\vec k_1 + \vec k_2}^2} (\omega_1 \omega_2 + \dotp{k_1}{k_2} - m^2) \right\} \\
 & + \frac{A_4} \alpha \sum_\pi (\omega_1 \omega_2 + \dotp{k_1}{k_2} - m^2 ) \Bigg\{ f(\vec k_3) \left[ \abs {k_3}^2 - \vec k_4 \cdot (\vec k_1 + \vec k_2) \frac{\abs{k_3}^2}{\mod{\vec k_1 + \vec k_2}^2} - \dotp{k_3}{k_4} \right] \\
& \qquad \qquad \qquad \qquad \qquad \qquad + f(\vec k_3) f(\vec k_4) \Bigg[ 2 \frac{\abs{k_3}^2 \abs{k_4}^2}{\mod{\vec k_1 + \vec k_2}^2} + 4 \abs{k_3}^2 \\ & \qquad \qquad \qquad \qquad \qquad \qquad \qquad \qquad \qquad + 4 \vec k_4 \cdot (\vec k_1 + \vec k_2)  \frac{\abs{k_3}^2}{\mod{\vec k_1 + \vec k_2}^2} + 2 \dotp{k_3}{k_4} \Bigg] \Bigg\} \\
& + \frac{B_4} {2 \alpha} \sum_\pi (\omega_1 \omega_2 + \dotp{k_1}{k_2} - m^2 ) \Bigg\{ f(\vec k_3) \Bigg[ - \abs{k_3}^4 - \abs{k_3}^2 \mod{\vec k_1 + \vec k_2}^2 + \dotp{k_3}{k_4} \\
& \qquad \qquad \qquad \qquad  \qquad \qquad \qquad \qquad  + \vec k_4 \cdot (\vec k_1 + \vec k_2) \frac{\abs{k_3}^2}{\mod{\vec k_1 + \vec k_2}^2} + \vec k_4 \cdot (\vec k_1 + \vec k_2) \abs{k_3}^2  \\
& \qquad \qquad \qquad \qquad  \qquad \qquad \qquad \qquad + \dotp{k_3}{k_4} \abs{k_3}^2 - 2 \abs{k_3}^2 \mod{\vec k_4 + \vec k_1 + \vec k_2}^2 \\
&  \qquad \qquad \qquad \qquad  \qquad \qquad \qquad \qquad - 2 \mod{\vec k_3 + \vec k_4}^2 \vec{k_3}^2 + \mod{\vec k_3 + \vec k_4}^2 \dotp{k_3}{k_4} \\
&  \qquad \qquad \qquad \qquad  \qquad \qquad \qquad \qquad + \abs{k_3}^2 \frac{\mod{\vec k_1 + \vec k_2 + \vec k_4}^2}{\mod{\vec k_1 + \vec k_2}^2} \vec k_4 \cdot (\vec k_1 + \vec k_2) \Bigg] \\ 
& \qquad \qquad \qquad  \qquad \qquad + 2 f(\vec k_3) f(\vec k_4) \Bigg[ 4 \vec k_3 \cdot (\vec k_1 + \vec k_2) \frac{\abs {k_4}^2}{\mod{\vec k_1 + \vec k_2}^2} + 2 \dotp{k_3}{k_4} \mod{\vec k_1 + \vec k_2}^2  \\
&  \qquad \qquad \qquad \qquad  \qquad \qquad \qquad \qquad + 2 \mod{\vec k_3 + \vec k_4}^2 \abs{k_4}^2 + 2 \abs{k_3}^2 \abs{k_4}^2 \frac{\mod{\vec k_1 + \vec k_2 + \vec k_3}^2}{\mod{\vec k_1 + \vec k_2}^2} \\
&   \qquad \qquad \qquad \qquad  \qquad \qquad \qquad \qquad  + 2 \abs{k_4}^2 \mod{\vec k_1 + \vec k_2 + \vec k_4}^2 + 2 \mod{\vec k_3 + \vec k_4}^2 \dotp{k_3}{k_4} \\
&   \qquad \qquad \qquad \qquad  \qquad \qquad \qquad \qquad  + 4 \abs{k_4}^2 \vec k_3 \cdot (\vec k_1 + \vec k_2) \frac{\mod{\vec k_1 + \vec k_2 + \vec k_3}^2}{\mod{\vec k_1 + \vec k_2}^2} \Bigg] \Bigg\} \\
& - \frac 2 \alpha \frac{(1 - 3 \lambda)^2}{1 - \lambda}  (\omega_1 \omega_2 + \dotp{k_1}{k_2} - m^2) \frac {\omega_3 \omega_4} {\mod{\vec k_1 + \vec k_2}^2} \left\{ 1 - \frac 1 {1 - \lambda}  \left[ \frac{(\dotp{k_3}{k_4})^2}{\abs{k_3}^2 \abs{k_4}^2} - \lambda \right] \right\} \\
& - 8 \frac{1 - 3 \lambda}{(1 - \lambda)^2} \sum_\pi f(\vec k_4) \omega_1 \omega_3 \vec k_2 \cdot (\vec k_1 + \vec k_2) \left[ \frac{\left( (\vec k_1 + \vec k_2) \cdot \vec k_3 \right)^2}{{\abs k_3}^2 \mod{\vec k_1 + \vec k_2}^4} - \frac{\lambda}{\mod{\vec k_1 + \vec k_2}^2} \right] \\
& + 4 \frac{1-3\lambda}{1 - \lambda} \sum_\pi f(\vec k_3) \frac{\omega_1 \omega_4}{\mod{\vec k_1 + \vec k_2}^2} \vec k_2 \cdot (\vec k_1 + \vec k_2) \\
& - 8 \frac{1 - 3 \lambda}{1 - \lambda} \sum_\pi f(\vec k_4) \vec k_3 \cdot (\vec k_1 + \vec k_2) \frac{\pi_{ij}^{\vec k_1 + \vec k_2} {\vec k_2}_{i} {\vec k_3}_j \omega_1 \omega_3}{\abs{k_3}^2 \mod{\vec k_1 + \vec k^2}^2} \\
& + 2 \frac{1 - 3 \lambda}{1 - \lambda} \frac{\omega_1 \omega_3}{\abs {k_3}^2} \frac{\pi_{ij}^{\vec k_1 + \vec k_2}}{\mod{\vec k_1 + \vec k_2}^2} \Bigg\{ {\vec k_3}_i \left[ (\vec k_3 + \vec k_4) \cdot (\vec k_1 + \vec k_2) + 2 \dotp{k_3}{k_4} \right]
 \\ & \qquad \qquad \qquad \qquad  \qquad \qquad  
 + {\vec k_4}_i \left[ \vec k_3 \cdot (\vec k_1 + \vec k_2) - (1 - \lambda) \abs{k_3}^2 \right] \Bigg\} \\
 & + \frac{1 - 3 \lambda}{(1 - \lambda)^2} \sum_\pi \frac{\omega_1 \omega_3}{\mod{\vec k_1 + \vec k_2}^2} \vec k_2 \cdot (\vec k_1 + \vec k_2) \Bigg\{ 2 \frac{[(\vec k_1 + \vec k_2) \cdot \vec k_3)]^2}{\mod{\vec k_1 + \vec k_2}^2 \abs{k_3}^2} - 2 \lambda + 4 \frac{\vec k_3 \cdot (\vec k_1 + \vec k_2) \vec k_4 \cdot (\vec k_1 + \vec k_2)}{\mod{\vec k_1 + \vec k_2}^2 \abs{k_3}^2} \\
 & \qquad \qquad \qquad \qquad \qquad \qquad + \frac{4 (\dotp{k_3}{k_4}) \vec k_3 \cdot (\vec k_1 + \vec k_2)}{\mod{\vec k_1 + \vec k_2}^2 \abs{k_3}^2} - 2 (1 - \lambda) \frac{\dotp{k_3}{k_4}}{\abs {k_3}^2} - 2 (1- \lambda) \frac{\vec k_4 \cdot (\vec k_1 + \vec k_2)}{\mod{\vec k_1 + \vec k_2}^2} \Bigg\}
\end{split}
\eeq
\normalsize
where $\sum_\pi$ means you permute over $\{1,2\}\{3,4\}$ and
\begin{subequations}
\begin{align}
\langle H_2 \rangle &= - \frac{1 - 3 \lambda}{1 - \lambda} \omega_3 \omega_4 \left[ 1 + 3 \lambda + \frac{1 - 3 \lambda}{1 - \lambda} \left( \frac{(\dotp{k_3}{k_4})^2}{\abs{k_3}^2 \abs{k_4}^2} - \lambda \right) \right] + \frac 1 2 \sum_{3 \leftrightarrow 4} ( 4 \abs{k_3}^2 + 6 \dotp{k_3}{k_4} ) \nonumber \\
& \qquad + 4 \alpha \left[ f(\vec k_3) \left( \abs{k_3}^2 + \dotp{k_3}{k_4} \right) + f(\vec k_3) f(\vec k_4) \left( 2 \abs{k_3}^2 - \dotp{k_3}{k_4} \right) \right] \nonumber \\
& \qquad + \frac 1 2 \sum_{3 \leftrightarrow 4} \Bigg\{  - \abs{k_3}^2 \abs{k_4}^2 (5 A_1 + 16 A_2) - A_1 (\dotp{k_3}{k_4})^2 \nonumber \\
& \qquad \qquad \qquad + A_3 \Big[ 4 \abs{k_3}^2  \left( \abs{k_4}^2 + \dotp{k_3}{k_4} \right) - 8 f(\vec k_3) \left( \abs{k_4}^4 + \abs{k_3}^2 \abs{k_4}^2 \right) \nonumber \\
& \qquad \qquad \qquad \qquad  + \mod{\vec k_3 + \vec k_4}^2 \left( - 6 \dotp{k_3}{k_4} - 16 \abs{k_3}^2 - 8 f(\vec k_3) \abs{k_4}^2 \right) \Big] \nonumber \\
& \qquad \qquad \qquad + A_4 \Big[ 4 f(\vec k_3) \abs{k_3}^2 \left( \dotp{k_3}{k_4} - \abs{k_3}^2 \right) + 4 f(\vec k_3) f(\vec k_4) \left( 2 \abs{k_3}^4 - \abs{k_3}^2 \abs{k_4}^2 \right) \nonumber \\
& \qquad \qquad \qquad \qquad + 4 f(\vec k_3) \mod{\vec k_3 + \vec k_4}^2 \left( \dotp{k_3}{k_4} - 2 \abs{k_3}^2 \right) + 8 f(\vec k_3) f(\vec k_4) \mod{\vec k_3 + \vec k_4}^2 \dotp{k_3}{k_4} \Big] \Bigg\}  \\
& \qquad + \frac 1 2 \sum_{3 \leftrightarrow 4} \Bigg\{ \dotp{k_3}{k_4} \Big[ \abs{k_3}^2 \abs{k_4}^2 (5 B_1 + 16 B_2) + B_1 (\dotp{k_3}{k4})^2 \Big] \nonumber \\
& \qquad \qquad \qquad + B_3 \Big[ - 12 \abs{k_4}^6 - 4 \dotp{k_3}{k_4} \abs{k_4}^4 + 8 f(\vec k_3) (\abs{k_4}^4 \abs{k_3}^2 - \abs{k_4}^6)   \nonumber \\
& \qquad \qquad \qquad \qquad + \mod{\vec k_3 + \vec k_4}^2 \left( - 8 f(\vec k_3) \abs{k_4}^4 + 8 \abs{k_4}^4 - 4 \dotp{k_3}{k_4} \abs{k_4}^2 \right)  \nonumber \\
& \qquad \qquad \qquad \qquad + \mod{\vec k_3 + \vec k_4}^4 \left( 6 \dotp{k_3}{k_4} + 16 \abs{k_4}^2 \right) \Big] \nonumber \\
& \qquad \qquad \qquad + B_4 \Big[ 4 f(\vec k_3) ( 3 \abs{k_3}^6 - \dotp{k_3}{k_4} \abs{k_3}^4) + 4 f(\vec k_3) f(\vec k_4) (\abs{k_3}^2 \abs{k_4}^4 - 2 \abs{k_4}^6) \nonumber \\
& \qquad \qquad \qquad \qquad + 2 f(\vec k_3) \mod{\vec k_3 + \vec k_4}^2 (8 \abs{k_3}^4 - 2 \dotp{k_3}{k_4} \abs{k_3}^2) \nonumber \\
& \qquad \qquad \qquad \qquad + 2 f(\vec k_3) \mod{\vec k_3 + \vec k_4}^4 (4 \abs{k_3}^2 - 2 \dotp{k_3}{k_4}) \nonumber \\
& \qquad \qquad \qquad \qquad + 8 f(\vec k_3) f(\vec k_4) \mod{\vec k_3 + \vec k_4}^4 (- \abs{k_4}^2 - \dotp{k_3}{k_4}) \Big] \Bigg\}
   \nonumber \\
\langle P_2 \rangle &= \half \sum_{3 \leftrightarrow 4} i \omega_4 \Bigg \{ \frac{1 - 3 \lambda}{1 - \lambda} f (\vec k_3) \Big[-2   \mod{\vec k_3 + \vec k_4}^2 + 2 \left( (\vec k_3 + \vec k_4) \cdot \vec k_4 \right)^2 \Big] \nonumber \\
& \qquad \qquad + \frac{1 - 3 \lambda}{1 - \lambda} \frac 1 {\abs {k_4}^2} \Big[ - 2 (\vec k_3 + \vec k_4) \cdot \vec k_3 (\vec k_3 + \vec k_4) \cdot \vec k_4 + (1 - \lambda) \mod{\vec k_3 + \vec k_4}^2 \dotp{k_3}{k_4} \Big] \nonumber \\ 
& \qquad \qquad + (2 \vec k_3 - (1-9 \lambda) \vec k_4) \cdot (\vec k_3 + \vec k_4) + \frac{1 - 3\lambda}{1 - \lambda} (\vec k_3 + \vec k_4) \cdot (\vec k_4 - 3 \vec k_3 )  \\ & \qquad  \qquad \qquad \qquad - 2 \lambda \left( 3 - \frac{1 - 3 \lambda}{1 - \lambda} \right) \mod{\vec k_3 + \vec k_4}^2 \Bigg\} \nonumber \\
\langle Q_2 \rangle_i &= \half \sum_{3 \leftrightarrow 4} \omega_4 \pi_{ij}^{\vec k_3 + \vec k_4} \Bigg\{ {\vec k_3}_j \left[ \frac{1- 3 \lambda}{1 - \lambda} \frac{\vec k_4 \cdot (\vec k_3 + \vec k_4)}{\abs{k_4}^2} - 2 + 3 \frac{1- 3 \lambda}{1 - \lambda} \right] \nonumber \\
& \qquad \qquad \qquad + {\vec k_4}_{j} \left[ 2 f(\vec k_3) \vec k_4 \cdot (\vec k_3 + \vec k_4) + \frac{1- 3 \lambda}{1 - \lambda} \vec k_3 \cdot (\vec k_3 + \vec k_4) + (1 - 9 \lambda) - \frac{1- 3 \lambda}{1 - \lambda} \frac{\dotp{k_3}{k_4}}{\abs{k_4}^2} \right] \Bigg\}
\end{align}
\end{subequations}

\end{document}